\documentclass[english]{article}
\usepackage[T1]{fontenc}
\usepackage{amssymb}
\usepackage{graphicx}

\makeatletter
\usepackage{a4}
\input{epsfig.sty}
\def\fnum@table{\tablename~{\bf\thetable}}
\def\fnum@figure{\figurename~{\bf\thefigure}}
\def\tablename{\footnotesize{\bf Table}}
\def\figurename{\footnotesize{\bf Figure}}

\def\be{\begin{equation}}
\def\ee{\end{equation}}

\makeatother

\usepackage{babel}
\begin{document}

\title{\textbf{A basic model for high energy cosmic ray interactions}}

\author{Sergey Ostapchenko$^{1}$,  Tanguy Pierog$^{2}$
  and G\"unter Sigl$^{1}$\\
$^1$\textit{\small Universit\"at Hamburg, II Institut f\"ur Theoretische
Physik, 22761 Hamburg, Germany}\\
$^2$\textit{\small  Institute for Astroparticle Physics, Karlsruher Institut für Technologie,}\\
\textit{\small Hermann-von-Helmholtz-Platz 1, 76344 Eggenstein-Leopoldshafen, Germany}\\
}

\maketitle
\begin{center}
\textbf{Abstract}
\par\end{center}
A    Monte Carlo generator of high energy cosmic ray interactions, 
relying on a very basic and transparent theoretical formalism,
in the framework of the Reggeon Field Theory,  is presented. The main motivation for our work
is to provide a new  cosmic ray interaction model characterized by relatively transparent physics, sufficient parameter freedom, and a high computational efficiency, which can be easily managed by external users, including a re-tuning of the model parameters. Such a model can be used for studying potential modifications of the interaction treatment, necessary for describing particular sets of data on extensive air showers  initiated by high energy cosmic rays, at a microscopic level, thereby keeping a consistency with general restrictions, like the unitarity, energy-momentum and charge conservation, Lorentz and isospin invariance. Importantly, this should allow one to study a compatibility of such modifications with relevant accelerator data. The  model  results for particle production and for 
basic extensive air shower  characteristics are presented and discussed.

\section{Introduction\label{intro.sec}}
Studies of ultra high energy cosmic rays (UHECRs) are traditionally performed using indirect
detection techniques: inferring the properties of primary cosmic ray (CR) particles from
measured characteristics of so-called
 extensive air showers (EAS) -- nuclear-electromagnetic cascades 
initiated by interactions of primary CRs in the atmosphere of the Earth \cite{nag00,blu09}.
 Consequently, such experimental activities imply an extensive use of  numerical
modeling of EAS development. A special role in   EAS simulations is played by Monte Carlo
(MC) generators of hadronic interactions, which form a bridge from accelerator experiments
studying such interactions in great detail to the CR field and which are responsible for %the quality of
 the description of EAS backbone -- the cascade of nuclear interactions of both
primary CRs and of produced secondary hadrons in the atmosphere \cite{eng11}.  MC generators
of CR interactions have been  considerably improved  over the past %three 
decades,
reaching a high level of sophistication \cite{pie15,rie20,ost24a}, while the parameters
of those models have been tuned based on a variety of accelerator data.

However, recent UHECR studies revealed that a number of experimental observations can not be
consistently described using present CR interaction models \cite{abr13,aab14,aab15,aab16}.
  Because of their considerable complexity, such models are typically regarded by
experimental groups as ``black boxes'', since a retuning of model parameters
 by external users has not generally  been envisaged.
This motivated one to consider an {\em ad hoc} rescaling of the predictions of EAS
simulations for particular air shower observables (e.g., \cite{abd24,ebr23}):  
to approach the  experimental observations. 
 While providing useful hints
 on how the simulations should be changed in order to better describe the EAS data,
 such a practice
 is potentially dangerous for a number of reasons.
 % suffer a lack of consistency.
 First of all, doing so one neglects correlations
between different EAS observables, which follow from a microscopic description of the 
interaction process.
Secondly and perhaps more importantly, proceeding
that way one is unable to check whether the desirable modifications 
are sensible in view
of existing ample experimental data on the properties of hadron-proton and hadron-nucleus
(nucleus-nucleus) collisions at high energies.
 Moreover, certain extreme modifications may 
 come into conflict with basic conservation laws.

All of the above motivated us to  develop  a new relatively simple MC generator of
hadronic interactions, staying within a very basic interaction picture,
 while using a reasonable and transparent formalism,
within the Reggeon Field Theory (RFT) approach \cite{gri68,gri69}.
Physics-wise, we are essentially coming back to the concepts of
the original Quark-Gluon String and Dual Parton models \cite{kai82,cap91}.
Therefore we use the name QGSb [Quark-Gluon Strings (basic)] for our model.
In more detail, we employ a phenomenological formalism, describing hadronic and nuclear scattering as mediated by Pomeron exchanges. We consider contributions to the scattering amplitude from two kinds of Pomerons -- corresponding to the underlying soft and semihard parton cascades. To extend the range of applicability of the model towards low energies, we take into consideration secondary Reggeon exchanges, using the ``effective Reggeon'' approach \cite{don92}. Production of secondary hadrons is treated as arising from fragmentation of strings of color field, stretched between consitituent partons of interacting
hadrons (nuclei).

It is worth stressing that we neither employ any miscroscopic treatment for parton cascades nor consider explicitly any nonlinear corrections. Rather,
we assume that the effects of both are effectively accounted for in our two-component Pomeron amplitude, in what  concerns  predictions for EAS characteristics. However, this restricts the range of applicability of the model. Without an explicit treatment of perturbative parton cascades, one is unable to describe correctly the production of hadrons at large transverse momenta. On the other hand, without an explicit treatment of nonlinear interaction effects, the model is inapplicable for treating heavy ion collisions. While both limitations are insignificant for EAS applications, they pose restrictions on accelerator data sets which can be used for the calibration of the model.

To some extent, the proposed model is complementary to existing more sophisticated MC generators of CR interactions. Restraining from using advanced theoretical approaches regarding the treatment of various aspects of hadronic collisions, we are unable to take into consideration certain important constraints, most notably, ones arising from experimental measurements of structure functions of hadrons. Such a strategy is intentional since we aim at a higher flexibility of the model, compared to more advanced MC generators which are typically overconstrained, within a particular theoretically advanced but yet phenomenological approach.

The main motivation for our work is twofold. Firstly, we wish to have a model
characterized by a sufficient parameter freedom, constrained mainly by accelerator data,
as a basic framework for studying uncertainties for  EAS predictions.
Secondly and more importantly, we want to provide experimental teams with a MC generator
characterized by relatively transparent physics and a sufficient parameter freedom,
which can be used for studying potential modifications of the interaction treatment,
necessary for describing particular sets of EAS data, at a microscopic level, thereby
keeping a consistency with general restrictions, like the unitarity, energy-momentum
and charge  conservation, Lorentz and isospin invariance. The goal is to allow the
experimentalists to check the ``cost'' of describing particular EAS data sets, in 
terms of (in)consistency with accelerator data.

Here we present the first version of the QGSb MC generator,
designed to treat hadronic interactions
in a broad energy range: from $\sim 10$ GeV laboratory (lab.)   energy
up to the highest energies of cosmic rays observed ($\sim 10^{11}$ GeV lab.). 
In the cause of its development, we  used and suitably
adapted various  relevant 
 technical procedures of the QGSJET(-II/III) models
\cite{kal93,ost11,ost24b}.
% e.g., regarding calculations of  hadronic 
%  interaction cross sections, generation of nucleon distributions in nuclei, 
% energy-momentum partition between multiple scattering processes, string fragmentation, 
% production of nuclear fragments, etc.
The outline of the paper is as follows. In Section \ref{physics.sec}, we discuss the
basic physics underlying our approach. In Section \ref{xsections.sec}, the expressions
for   hadronic interaction cross sections are provided,
while Section \ref{prod.sec} is devoted to a description of the treatment of secondary particle 
production. The results of the model for various characteristics of hadron-proton
and hadron-nucleus %(nucleus-nucleus) 
collisions are presented in Section \ref{results.sec}, in comparison
to accelerator data. In Section \ref{eas.sec}, the energy dependence of basic EAS
characteristics, predicted by the model, is presented and discussed. Finally we
conclude in Section \ref{concl.sec}.

\section{Basic physics  \label{physics.sec}}

High energy hadronic collisions are predominantly multiple scattering processes, 
being mediated by (generally nonperturbative)  parton cascades 
developing between the interacting projectile
and target hadrons (nuclei). For such parton cascades, we  use an effective
macroscopic description -- treating them as Pomeron exchanges. Here we have to take
into account two mechanisms which impact the ``parton content'' of the Pomeron.
First, with increasing energy, the contribution of  the so-called 
semihard processes corresponding to parton cascades developing, at least partly,
in the  domain of high parton virtualities $q^2$ becomes  important --
since the smallness of the corresponding strong coupling, $\alpha_s(q^2)$, in such
cascades becomes compensated by large collinear and infrared logarithms and by a
high parton density \cite{glr}. 
 Instead of treating such perturbative cascades 
explicitly, we   consider two contributions
to the Pomeron amplitude: the one corresponding to ``soft'' (low  virtuality)
parton cascade, characterized by a rather slow energy rise, and the ``semihard''
one quickly rising with energy \cite{don98,ost02}. Further, at sufficiently high energies,
an important role in the interaction dynamics is played by nonlinear effects arising
from ``splitting'' and ``fusion'' of parton cascades, both real and virtual ones,
typically described as  Pomeron-Pomeron interactions (e.g., \cite{ost06}).
Here we assume that the main consequence of such nonlinear corrections is to
``renormalize'' the Pomeron amplitude \cite{kai86}, yielding, in particular,
a small intercept of the ``soft'' Pomeron component, corresponding to the 
above-mentioned slow energy rise of the corresponding contribution.
 Thus, we expect that our two-component Pomeron amplitude effectively takes
  into account the effects
of both mechanisms: the perturbative parton cascading and the nonlinear corrections 
to the interaction dynamics, in what concerns the predictions for EAS characteristics.

In order to extend the model for a description of hadronic interactions at relatively low energies, below 100 GeV lab.,
 one has to account for secondary Reggeon exchanges,
in addition to the Pomeron  contributions (e.g., \cite{col77}).
In that respect, we follow the approach proposed in  \cite{don92}:
 considering an exchange of a single effective Reggeon per reaction type,
 rather than treating explicitly all the relevant Reggeon contributions.

When describing high energy hadronic collisions, it is important to
take into consideration
inelastic diffraction processes and the   closely related inelastic screening 
 (``color fluctuation'') effects \cite{gri69,fra08}.
To treat low mass diffractive excitations of interacting hadrons (nucleons),
we   rely on the Good-Walker  (GW) approach \cite{goo60}:  considering hadrons 
to be represented by superpositions of a large number
of Fock states characterized by different transverse sizes and by different
couplings to the Pomerons, which diagonalize the scattering matrix,
assuming a coupling to be proportional to the
transverse area of the state. It is noteworthy that this mechanism gives rise to three
important effects: a milder energy rise of the interaction cross sections,
compared to the case when such inelastic screening corrections are neglected, 
 especially, for hadron-nucleus and nucleus-nucleus 
collisions, the emergence of low mass diffraction, and larger fluctuations
for secondary particle production.
Regarding a treatment of high mass diffraction, a
 self-consistent approach implies an all-order resummation of enhanced Pomeron 
 diagrams \cite{ost10}. Here we   rely instead   on an effective description: 
 considering only the diffractive cut of the simplest triple-Pomeron  graph 
 corresponding to an interaction between three ``soft'' Pomerons, as described in  Section \ref{prod.sec}.

Having defined the Pomeron exchange amplitude,
 including  the Pomeron coupling to various Fock states, 
 one  can calculate total, elastic, and inelastic
cross sections for hadron-proton, hadron-nucleus, and nucleus-nucleus collisions.
Moreover, considering unitarity cuts of the corresponding elastic scattering diagrams
and applying the so-called Abramovskii-Gribov-Kancheli %(AGK)
cutting rules \cite{agk74}, one is able to obtain partial cross sections for various
inelastic final states corresponding to having precisely $n$  ``elementary'' production
processes ($n$ ``cut Pomerons''), hence, to sample ``macro-configurations'' of the 
collisions, using  MC methods. However, for treating secondary particle
production, one has to define the hadronization procedure. In that respect, we  
traditionally assume that each cut Pomeron   corresponds
to a creation of a pair of strings of color field, stretched between constituent partons
[(anti)quarks or (anti)diquarks] of the interacting hadrons \cite{kai82,cap91}.
 With such partons flying apart,
the strings break up and hadronize via a creation of quark-antiquark and diquark-antidiquark
pairs from the vacuum. Here we employ the string fragmentation procedure of the 
QGSJET model \cite{kal93}, the  current implementation being described in  \cite{ost24b}.
While the parameters of this procedure can be expressed via intercepts of relevant
Regge trajectories \cite{kai87},  we  shall generally
  treat them as adjustable ones. %parameters,
%when fitting accelerator data.
 In particular, since one  expects important differences between soft and semihard parton
cascades, notably, a higher parton density per unit rapidity for the latter, the string
fragmentation parameters will generally differ between the  soft and semihard Pomerons.
  
  Apart from secondary hadron production, of importance for extensive air shower simulations is a treatment of the fragmentation of the nuclear spectator part
  composed of all noninteracting nucleons, in nucleus-nucleus collisions:
  since this   impacts fluctuations of nucleus-induced EAS \cite{kal93,eng91}.
Such a nuclear breakup is modeled using  the percolation-evaporation approach of the QGSJET model \cite{kal93}.

\section{Cross section formulas  \label{xsections.sec}}

\subsection{High energy collisions  \label{xsections-high.sec}}

The Pomeron exchange eikonal (the imaginary part\footnote{The real part of the Pomeron
amplitude can be neglected in the high energy limit.}
 of the corresponding amplitude in the impact parameter representation)
  for hadron-proton collisions is usually chosen in the form
\begin{equation}
\chi_{hp}^{\mathbb P}(s,b)=\frac{\gamma_h\,\gamma_p\,
(s/s_0)^{\alpha_{\mathbb P}(0)-1}}{R_h^2+R_p^2+
\alpha_{\mathbb P}'\,\ln (s/s_0)}\;
 \exp \!\left[- \frac{b^2/4}{R_h^2+R_p^2+\alpha_{\mathbb P}'\,\ln (s/s_0)}\right],
\label{eq:chi-pom}
\end{equation}
where $\alpha_{\mathbb P}(0)$ and $\alpha_{\mathbb P}'$ are, respectively, the intercept
and the slope of the Pomeron Regge trajectory, $\gamma_h$ is the residue and $R_h^2$ is the
slope for   Pomeron coupling to hadron $h$, and $s_0 \simeq 1$ GeV$^2$ -- the hadronic
mass scale; $s$ and $b$ are, respectively,  the center-of-mass (c.m.) energy squared 
and the impact parameter for the collision.

As discussed  in Section \ref{physics.sec},  we shall consider contributions to $\chi_{hp}^{\mathbb P}$ from  soft
and semihard Pomerons, characterized by different energy dependence, i.e., by
different overcriticalities (intercept minus unity) $\Delta_{\rm s}$ and  $\Delta_{\rm sh}$,
and by different couplings to hadrons.
We also assume hadron states to be composed of a number of GW components:
\begin{equation}
|h\rangle=\sum_{i=1}^{N_{\rm GW}}\sqrt{C^{(i)}_{h}}\,|i\rangle \label{eq:gw-h}, \label{eq:gw-h*}
\end{equation}
where $C^{(i)}_{h}$  are the corresponding partial weights
($\sum_{i}C^{(i)}_{h}=1$) and $N_{\rm GW}$ is the total number of GW states considered per hadron.

Thus,  a Pomeron exchange between Fock states $|i\rangle$ and  $|j\rangle$
of the projectile hadron and of the target proton, respectively, will be described
by the eikonal
\begin{equation}
\chi_{hp(ij)}^{\mathbb P}(s,b)=\frac{\gamma_{h(i)}\,\gamma_{p(j)}\,
(\delta_{\rm s/sh}\,s^{\Delta_{\rm s}}+(1-\delta_{\rm s/sh})\,s^{\Delta_{\rm sh}})}{R_{h(i)}^2+R_{p(j)}^2+
\alpha_{\mathbb P}'\,\ln s}\;
 \exp \!\left[- \frac{b^2/4}{R_{h(i)}^2+R_{p(j)}^2+
\alpha_{\mathbb P}'\,\ln s}\right].
\label{eq:chi-pom-gen}
\end{equation}
Here $\gamma_{h(i)}$ and $R_{h(i)}^2$ are the residue and  the
slope for  Pomeron coupling to Fock state  $|i\rangle$ of hadron $h$, while 
$\delta_{\rm  s/sh}$ characterizes the relative weights of the soft and semihard
 Pomerons, which is related to a weaker coupling of the latter  to hadrons.
 This is assumed
to be universal for all hadron types and for all Fock states. Further, we use here
the same slope $\alpha_{\mathbb P}'$ for the   soft
and  semihard Pomerons, i.e., we assume the same rate of transverse expansion for the
corresponding underlying parton cascades. Generally, this might not be the case
since a semihard   cascade develops partly in a high parton virtuality $q^2$ domain,
where parton transverse displacements $\Delta b^2 \propto 1/q^2$ are small, which may
result in a smaller Pomeron slope.

Using the eikonals (\ref{eq:chi-pom-gen}),   one obtains  for the total and 
elastic hadron-proton cross sections:
\begin{eqnarray}
\sigma^{\rm tot}_{hp}(s) 
=  2\int \! d^2b \;\sum_{i,j}C^{(i)}_{h}C^{(j)}_{p} 
\left[1-e^{-\chi_{hp(ij)}^{\mathbb P}(s,b)}\right]
\label{eq:sig-eik-tot-gw} &&\\
\sigma^{\rm el}_{hp}(s) =   \int \! d^2b  \left[\sum_{i,j}C^{(i)}_{h}C^{(j)}_{p}\,
(1-e^{-\chi_{hp(ij)}^{\mathbb P}(s,b)})\right]^2.
\label{eq:sig-eik-el-gw}  && 
\end{eqnarray}

In turn, for the ``topological'' cross section corresponding to having precisely $n$
elementary production processes (cut Pomerons), one has
\begin{equation}
\sigma^{(n)}_{hp}(s) = \int \! d^2b \; \sum_{i,j}C^{(i)}_{h}C^{(j)}_{p}\, 
 \frac{\left[2\chi_{hp(ij)}^{\mathbb P}(s,b)\right]^n}{n!}\,
e^{-2\chi_{hp(ij)}^{\mathbb P}(s,b)}\,.
   \label{eq:sig(n)-gw} 
\end{equation}

The total inelastic cross section, 
$\sigma^{\rm inel}_{hp}=\sigma^{\rm tot}_{hp}-\sigma^{\rm el}_{hp}$,
consists of the ``absorptive'' part,
\begin{equation}
\sigma^{\rm abs}_{hp}(s) =\sum_{n=1}^{\infty} \sigma^{(n)}_{hp}(s) = \int \! d^2b 
\; \sum_{i,j}C^{(i)}_{h}C^{(j)}_{p}\, 
 \left[1-e^{-2\chi_{hp(ij)}^{\mathbb P}(s,b)}\right],
   \label{eq:sig-abs} 
\end{equation}
and of the cross section for low mass diffraction. The latter splits into the
projectile, target, and double diffraction cross sections:
\begin{eqnarray}
\sigma_{hp}^{{\rm SD(proj)}}(s)=\int\! d^{2}b\;\sum_{i,k,j,l}
(C^{(i)}_{h}\,\delta_{ik}-C^{(i)}_{h}C^{(k)}_{h})\, C^{(j)}_{p}C^{(l)}_{p}
\;  e^{-\chi_{hp(ij)}^{\mathbb P}(s,b)-\chi_{hp(kl)}^{\mathbb P}(s,b)} && 
\label{eq:sig-sd-proj}\\
\sigma_{hp}^{{\rm SD(targ)}}(s)=\int\! d^{2}b\;\sum_{i,k,j,l}C^{(i)}_{h}C^{(k)}_{h}\,
(C^{(j)}_{p}\,\delta_{jl}-C^{(j)}_{p}C^{(l)}_{p})\, \;  e^{-\chi_{hp(ij)}^{\mathbb P}(s,b)-\chi_{hp(kl)}^{\mathbb P}(s,b)} && 
\label{eq:sig-sd-targ}\\
\sigma_{hp}^{{\rm DD}}(s)=\int\! d^{2}b\;\sum_{i,k,j,l}
(C^{(i)}_{h}\,\delta_{ik}-C^{(i)}_{h}C^{(k)}_{h})\, (C^{(j)}_{p}\,\delta_{jl}-C^{(j)}_{p}C^{(l)}_{p})
\;  e^{-\chi_{hp(ij)}^{\mathbb P}(s,b)-\chi_{hp(kl)}^{\mathbb P}(s,b)}\,. && 
\label{eq:sig-dd}
\end{eqnarray}

A generalization of Eqs.\ (\ref{eq:sig-eik-tot-gw}-\ref{eq:sig-dd}) for  hadron-nucleus 
and nucleus-nucleus collisions can be found in \cite{kal93,zol88}.
In particular, for the total and particle production cross sections for hadron-nucleus interactions, one has
\begin{eqnarray}
\sigma^{\rm tot}_{hA}(s) 
&=&  2\int \! d^2b \;\sum_{i}C^{(i)}_{h}\left\{1-\left[\int \!d^2\kappa \;T_A(\kappa)\,\left(1-
\sum_{j}C^{(j)}_{p}\Gamma ^{(ij)}_{hp} (s,|\vec b-\vec \kappa |)\right)\right]^A\right\}
\label{eq:sigtot-ha} \\
\sigma^{\rm prod}_{hA}(s) 
&=&  \int \! d^2b \;\sum_{i,k}C^{(i)}_{h}C^{(k)}_{h}
\left\{1-\left[
\int \!d^2\kappa \;T_A(\kappa) \left(1-\sum_{j}C^{(j)}_{p}\Gamma ^{(ij)}_{hp}(s,|\vec b-\vec \kappa |)\right)\right.\right. \nonumber \\
&\times & \left.\left.
\left(1-\sum_{l}C^{(l)}_{p}\Gamma ^{(kl)}_{hp}(s,|\vec b-\vec \kappa |)\right)\right]^A\right\}.
\label{eq:sigin-ha}
\end{eqnarray}
Here $\Gamma ^{(ij)}_{hp}$ is the profile function for a scattering of Fock state
 $|i\rangle$ of the projectile hadron on Fock state  $|j\rangle$  of a target nucleon,
 \begin{equation}
 \Gamma ^{(ij)}_{hp}(s,b)=1- e^{-\chi^{\mathbb P}_{hp(ij)}(s,b)}, \label{eq:hn-prof}
\end{equation}
while $T_A$ is the nuclear profile expressed via the nuclear density $\rho_A$ as
 \begin{equation}
 T_A(r_{\perp})=\int \!dz\;\rho_A(\vec r)\,.
  \label{eq:nucl-prof}
\end{equation}

As discussed  in Section \ref{physics.sec},  we assume the Pomeron couplings to different Fock states of hadrons
to be proportional to the transverse radii squared of those states:
 \begin{equation}
\gamma_{h(i)}=g_0\,R_{h(i)}^2,\label{eq:gamma-r} 
\end{equation}
with  $g_0$ being thus (up to a factor)  the coupling per unit transverse area.
Further, to reduce the number of adjustable parameters, we consider equal weights for the
different Fock states, 
 \begin{equation}
  C^{(i)}_{h} \equiv 1/N_{\rm GW},
\end{equation}
  and choose a  loguniform distribution for $R^2_{h(i)}$:
\begin{eqnarray}
R^2_{h(i)}=R^2_{h(1)}\,d_h^{\frac{i-1}{N_{\rm GW}-1}}\label{eq:R_i}.
\end{eqnarray}
 
Thus, in the minimal scheme described above, a treatment of proton-proton and proton-nucleus
(nucleus-nucleus) collisions involves 7 adjustable parameters:  $\Delta_{\rm s}$, $\Delta_{\rm sh}$, $\alpha_{\mathbb P}'$,  $\delta_{\rm s/sh}$, $g_0$, $R^2_{p(1)}$, and
$d_p$,  which can be
fixed using experimental data on the total, elastic, and diffractive proton-proton cross sections.
Considering also interactions of pions and kaons, one has additionally  the parameters 
$R^2_{\pi(1)}$, $R^2_{K(1)}$, $d_{\pi}$, and $d_K$, whose values can be fixed based on
 measured pion-proton and kaon-proton cross sections.
It is worth discussing the basic correspondence between
these parameters and experimental observables.  The values of the
overcriticalities  $\Delta_{\rm s}$ and  $\Delta_{\rm sh}$ have the largest impact on the 
energy dependence of the total, elastic, and inelastic cross sections at ``intermediate''
and very high energies, respectively, while the parameter  $\delta_{\rm s/sh}$ controls the
energy where the transition between the two trends takes place.\footnote{See also the corresponding discussion in Section \ref{results.sec} and Fig.\ \ref{fig:sig-inel} there.}
 The Pomeron slope  $\alpha_{\mathbb P}'$ governs the
shrinkage of the diffractive cone, being largely constrained by measurements of the
proton elastic scattering slope,
 $B^{\rm el}_{pp}=d\ln \left.d\sigma^{\rm el}_{pp}/dt\right|_{t=0}$.
 The other parameters,  $g_0$, $R^2_{p(1)}$, and $d_p$, have a strong impact on the absolute
values of $\sigma^{\rm tot}_{pp}$, $\sigma^{\rm el}_{pp}$, and $B^{\rm el}_{pp}$, hence, 
can be roughly fixed based on fixed target data. Moreover, the parameter
 $d_p\equiv R^2_{p(N_{\rm GW})}/R^2_{p(1)}$  (similarly,
 $d_{\pi}$  and $d_K$ for pions and kaons) controls the differences between various Fock states
 of the proton and thereby governs the rate of low mass diffraction: the diffractive cross sections
(\ref{eq:sig-sd-proj}-\ref{eq:sig-dd}) rise with a decrease of  $d_p$ or vanish for  $d_p=1$. It is noteworthy that changing the number of GW states considered,
$N_{\rm GW}$, requires a retuning of all the above-discussed parameters.

\subsection{Extension to low energies\label{xsections-low.sec}}
In order to extend our treatment to low energies, below 100 GeV lab.,
 we have to take into consideration secondary Reggeon ($\rho$, $\omega$, $f$, $A_2$, where relevant)
  contributions, in addition to Pomeron
exchanges. The hadron-proton scattering amplitude thus becomes
\begin{equation}
A_{hp}(s,t)=2is\sum_{i,j}C^{(i)}_{h}C^{(j)}_{p} \int \!d^2b\; e^{i\vec q_{\perp}\vec b}
 \left[1-e^{-\chi_{hp(ij)}^{\mathbb P}(s,b)}\right]
+\sum _{{\mathbb R}_i} A_{hp}^{{\mathbb R}_i}(s,t)\,, \label{Eq: A(s,t)}
\end{equation}
where $t\simeq -q_{\perp}^2$ is the squared momentum transfer for the process and the Reggeon exchange amplitude 
$A_{hp}^{{\mathbb R}_i}$ is defined as (e.g., \cite{col77})
\begin{equation}
 A_{hp}^{{\mathbb R}_i}(s,t)= s_0\,\gamma_h^{{\mathbb R}_i}(t)\,\gamma_p^{{\mathbb R}_i}(t)\,
 \eta_{{\mathbb R}_i}(t)\,(s/s_0)^{\alpha_{{\mathbb R}_i}(t)}. \label{Eq: A-reg}
\end{equation}
Here $\alpha_{{\mathbb R}_i}(t)$ is the Regge trajectory  for $i$-th Reggeon, 
$\gamma_h^{{\mathbb R}_i}$ is its coupling to hadron $h$, and $\eta_{{\mathbb R}_i}$ is the
signature factor,
\begin{equation}
 \eta_{{\mathbb R}_i}(t)=- \frac{\sigma_{{\mathbb R}_i}+e^{-i\,\pi \,\alpha_{{\mathbb R}_i}(t)}}
 {\sin [\pi \, \alpha_{{\mathbb R}_i}(t)]}\,,
\end{equation}
with $\sigma_{{\mathbb R}_i}=\pm 1$.

A remarkable property of Regge trajectories is their linearity,
\begin{equation}
\alpha_{{\mathbb R}_i}(t)\simeq \alpha_{{\mathbb R}_i}(0)+\alpha '_{\mathbb R}\,t\,,
\end{equation}
with $\alpha '_{\mathbb R}\simeq 0.9$ GeV$^{-2}$. 
Moreover, there is a considerable
exchange degeneracy for the trajectories we are interested in:
\begin{equation}
\alpha_{\rho}(0)\simeq  \alpha_{\omega}(0)\simeq\alpha_{f}(0)\simeq\alpha_{A_2}(0) 
\equiv \alpha_{\mathbb R}\,, \label{Eq: generacy}
\end{equation}
with $\alpha_{\mathbb R}\simeq 0.5$, which allowed one to develop very successful
parametrizations of   total cross sections for hadron-proton scattering, considering
an exchange of a single effective Reggeon \cite{don92}.

Indeed,  using Eqs.\ (\ref{Eq: A(s,t)}-\ref{Eq: generacy}), we get
\begin{equation}
\sigma ^{\rm tot}_{hp}(s)=\frac{1}{s}{\rm Im}A_{hp}(s,0)
=2\sum_{i,j}C^{(i)}_{h}C^{(j)}_{p}  \int \!d^2b\;  \left[1-e^{-\chi_{hp(ij)}^{\mathbb P}(s,b)}\right]
+ \gamma_{hp}^{\mathbb R}\,(s/s_0)^{\alpha_{\mathbb R}-1},
\label{Eq: sig-tot}
\end{equation}
where $\gamma_{hp}^{\mathbb R}\equiv 
\sum _{{\mathbb R}_i}\gamma_h^{{\mathbb R}_i}(0)\,\gamma_p^{{\mathbb R}_i}(0)$
corresponds to the product of the residues for the effective Reggeon coupling
to hadron $h$ and the proton.

We shall go further and assume an exponential $t$-dependence for the vertex 
factors $\gamma_h^{{\mathbb R}_i}(t)$, using the same Regge
slopes $\lambda _h^{\mathbb R}$
for all the considered Reggeons, for a given hadron $h$:
\begin{equation}
 \gamma_h^{{\mathbb R}_i}(t)=\gamma_h^{{\mathbb R}_i}(0)\,e^{\lambda _h^{\mathbb R}t}.
 \label{Eq: reg-vertex}
\end{equation}

In addition, we take into account that   real parts of the amplitudes 
$A_{hp}^{{\mathbb R}_i}$ are largely canceled between the different Reggeons,
such that $|{\rm Re}A_{hp}(s,t)/{\rm Im}A_{hp}(s,t)|\lesssim 0.2$ 
in the energy range of our interest \cite{pdg}. This allows us to neglect the
real part of $A_{hp}$ since its contribution to the elastic cross section
$\sigma ^{\rm el}_{hp}$ is at the level of few per cent.

Thus, we finally obtain the scattering amplitude in the impact parameter representation as
\begin{eqnarray}
\tilde A_{hp}(s,b)&=&\frac{1}{8\pi^2s} \int \!d^2q_{\perp}\; e^{-i\vec q_{\perp}\vec b}\;A_{hp}(s,-q_{\perp}^2)
\nonumber \\
&=&i\sum _{i,j} C^{(i)}_h  C^{(j)}_p\left[1-e^{-\chi_{hp(ij)}^{\mathbb P}(s,b)}\right]
+ \frac{i\,\gamma_{hp}^{\mathbb R}\,(s/s_0)^{\alpha_{\mathbb R}-1}}
{8\pi\lambda _{hp}^{\mathbb R}}\,e^{-\frac{b^2}{4\lambda _{hp}^{\mathbb R}}}\,,
 \label{Eq: A(s,b)}
\end{eqnarray}
where $\lambda _{hp}^{\mathbb R}=\lambda _{h}^{\mathbb R}+\lambda _{p}^{\mathbb R}
+\alpha '_{\mathbb R}\,\ln (s/s_0)$. It is noteworthy that, when
 considering the Good-Walker decomposition of hadron wave functions,
we used here the same coupling of the effective Reggeon to all the GW states of a
 given hadron, thereby
neglecting color fluctuation effects in low energy scattering.

The total and elastic cross sections are given by the usual expressions:
\begin{eqnarray}
\sigma ^{\rm tot}_{hp}(s)=2 \int \!d^2b\; {\rm Im}\tilde A_{hp}(s,b)   \label{sigt-b} \\
\sigma ^{\rm el}_{hp}(s)= \int \!d^2b\; |\tilde A_{hp}(s,b)|^2 ,   \label{sigel-b}
\end{eqnarray}
with $\sigma ^{\rm inel}_{hp}=\sigma ^{\rm tot}_{hp}-\sigma ^{\rm el}_{hp}$.
In turn,  the total and particle production cross sections for hadron-nucleus scattering are 
given by Eqs.\ (\ref{eq:sigtot-ha}-\ref{eq:sigin-ha}), with the partial interaction profile
$\Gamma ^{(ij)}_{hp}$ being now defined as
 \begin{eqnarray}
 \Gamma ^{(ij)}_{hp}(s,b)&=&1- e^{-\chi^{\mathbb P}_{hp(ij)}(s,b)} +  \Gamma ^{\mathbb R}_{hp}(s,b)\, 
 \label{eq:hn-prof-low}\\
  \Gamma ^{\mathbb R}_{hp}(s,b) &=&  \frac{\gamma_{hp}^{\mathbb R}\,(s/s_0)^{\alpha_{\mathbb R}-1}}
{8\pi\lambda _{hp}^{\mathbb R}}\,e^{-\frac{b^2}{4\lambda _{hp}^{\mathbb R}}}\,.
 \label{eq:prof-rej}
\end{eqnarray}

\section{Treatment of particle production  \label{prod.sec}}
\subsection{Basic scheme  \label{prod-basic.sec}}
The procedure for modeling secondary hadron production follows the standard logic (see, e.g., 
\cite{ost24b}), further discussed for the case of hadron-proton interaction.
One  starts with sampling the impact parameter squared $b^2$ for the collision: uniformly
between 0 and a sufficiently large value $b^2_{\max}$,
and choosing the GW states  $|i\rangle$ and  $|j\rangle$
of the projectile hadron and of the target proton, respectively. 
Further, one samples an ``absorptive''
interaction, with the probability $w_{abs}$ defined by the integrand in the 
right-hand-side (rhs) of Eq.\ (\ref{eq:sig-abs}), or low mass 
diffractive processes, the corresponding probabilities 
being defined by the integrands in the rhs of 
Eqs.\ (\ref{eq:sig-sd-proj}-\ref{eq:sig-dd}).
If neither takes place, a sampling of the impact parameter and of the GW components
of the interacting hadrons is repeated.

For an absorptive interaction, one samples the number of cut Pomerons $n\geq 1$: according to the 
Poisson distribution with the average $2\chi_{hp(ij)}^{\mathbb P}(s,b)$ -- 
see the integrand in the rhs of Eq.\ (\ref{eq:sig(n)-gw}).  Further, one
proceeds with the energy-momentum sharing between $2n$ constituent partons of the projectile
hadron, to which strings of color field are attached, and similarly for $2n$ constituent
partons of the target proton. Here for the distribution of the  fractions $x^{\pm}$ of 
light cone (LC) momenta ($E\pm p_z$)  of  constituent partons (string ``ends''), we use
 \begin{equation}
 f_h^{\rm LC}(x_1^{\pm},...,x_{2n}^{\pm})\propto \left[\prod_{i=1}^{2n-2}
 (x_i^{\pm})^{-\alpha_{\rm sea}}  \right] 
  \; (x_{2n-1}^{\pm})^{-\alpha_{\mathbb{R}}}\,
 (x_{2n}^{\pm})^{-\alpha_{\rm r}}\, \delta \!\left(1-\sum_{j=1}^{2n}x_j^{\pm}\right) .
 \label{ems.eq}
 \end{equation}
 Here the exponent $\alpha_{\rm sea}$ is used for sea (anti)quarks, considering
 only  light $u$ and $d$ (anti)quarks as string ends, while the small $x$ limit
 of a valence (anti)quark distribution  follows the Regge behavior 
 ($\propto x_{2n-1}^{-\alpha_{\mathbb{R}}}$). 
 %with $\alpha_{\mathbb{R}}=0.5$.
 Similarly, the LC momentum distribution of the remaining valence parton
 [(anti)quark for $h$ being a meson or (anti)diquark for a (anti)baryon] follows the
 corresponding Regge behavior ($\propto x_{2n}^{-\alpha_{\rm r}}$) \cite{kai82,kai87}:
  \begin{eqnarray}
 \alpha_u  =\alpha_d  = \alpha_{\mathbb{R}} 
  \label{alpha-u.eq}\\
  \alpha_{ud}  =\alpha_{uu}+1  =2\alpha_{N}(0)- \alpha_{\mathbb{R}} 
 \label{alpha-ud.eq}
  \\ 
  \alpha_s  = \alpha_{\phi}(0)\\
  \alpha_{us}  =\alpha_{ds}  = \alpha_{ud}-\alpha_{\mathbb{R}} +  \alpha_{\phi}(0) \,.
 \label{alpha-us.eq}
   \end{eqnarray}
%   However, in the effective scheme we choose, we shall treat $\alpha_{\rm r}$ as
%    adjustable parameters.
 
Regarding the   LC momentum distribution of constituent sea quarks,
in principle, it should also be characterized by the Regge behavior
 ($\propto x_i^{-\alpha_{\mathbb{R}}}$) in the  low $x$ limit \cite{kai82},
 i.e., $\alpha_{\rm sea}=\alpha_{\mathbb{R}}$,
 which was adopted in the QGSJET(-II)   model and will also be used for
  the default tune of this model (see Section \ref{results.sec}).
However, generally $\alpha_{\rm sea}$ can be treated as an adjustable parameter,  
  $0.5 \leq \alpha_{\rm sea}<1$, since the energy-momentum sharing between
  constituent partons is expected to be modified by nonlinear interaction
  effects. Moreover, since such effects become stronger at higher energies,
  one may use different $\alpha_{\rm sea}$ values for string ends corresponding
  to  soft and  semihard Pomerons, namely, using a softer distribution
  (larger  $\alpha_{\rm sea}$) for the latter. In fact, it is the value of
  this parameter  which grossly defines the predictions for the 
  energy dependence of the inelasticity 
  of hadron-nucleus collisions, having thereby a decisive impact on the predicted
  EAS maximum depth $X_{\max}$ \cite{ost24c}.
    
 In turn,  transverse momenta of string ends are sampled according to the distribution:
   \begin{equation}
 f_h^{\perp}(\vec p_{t_1},...,\vec p_{t_{2n}})
 \propto 
 \exp\!\left(-\sum_{i=1}^{2n-1}\frac{p_{t_i}^2}{\gamma_{\perp}}
 -\frac{p_{t_{2n}}^2}{\gamma_{\perp}^h}\right) 
 \; \delta^{(2)}\!\left(1-\sum_{j=1}^{2n}\vec p_{t_j}\right).
\label{pts.eq}
 \end{equation}

\begin{figure*}[t]
\centering
\includegraphics[height=4cm,width=0.8\textwidth]{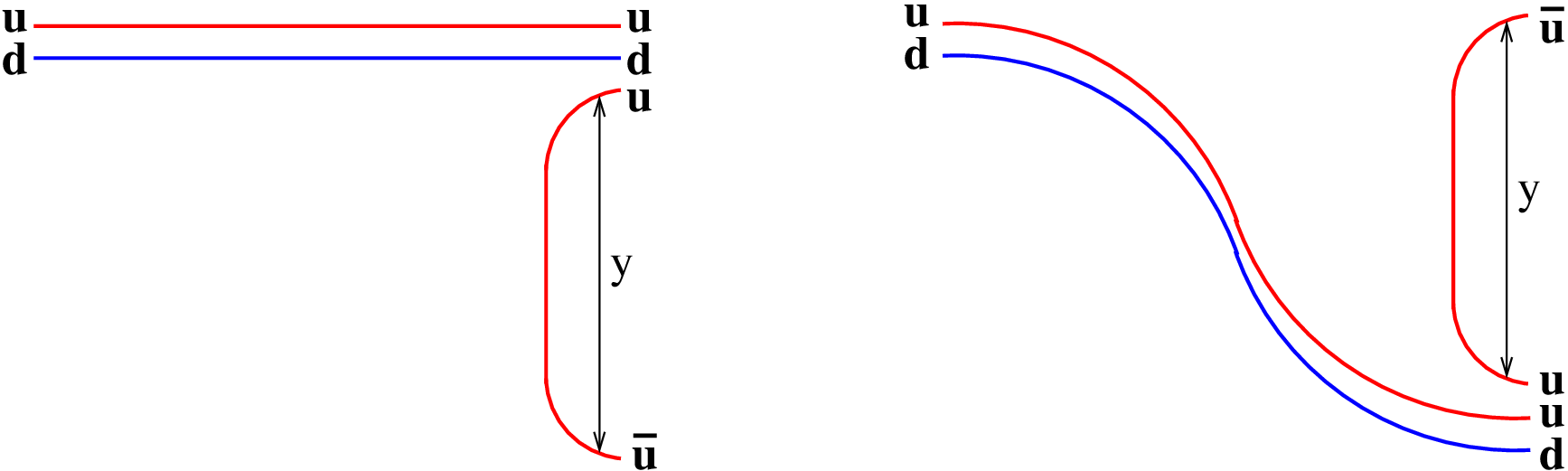}
\caption{Schematic view of the fragmentation of a fast $ud$ diquark into a proton:
both large $x$ and small $x$ limits of the proton LC momentum distribution
correspond to a large rapidity $y$ separation between the $u$ and $\bar u$
of the vacuum-created $u\bar u$ pair. Left: the large $x$ limit  
($1-x\propto e^{-y}$) is obtained  by slowing down the  $\bar u$ antiquark.
Right: the small  $x$ limit ($x\propto e^{-y}$) corresponds to slowing down 
both the $u$ quark and the original $ud$ diquark.}
\label{fig:fragm}       
\end{figure*}

The following step is the string fragmentation performed considering
iteratively an emission of a hadron from either string end, 
with the hadron LC momentum fraction $x$ (LC$^+$ or LC$^-$ for a projectile or
 target side end, respectively) and the transverse momentum
 $p_t$ in the c.m.\ frame of the string being sampled 
 according to the distribution:
   \begin{eqnarray}
 f_{qq'\rightarrow h}^{\rm fragm}(x,\vec p_{t}) 
 \propto x^{1-\alpha_q-\alpha_{q'}}(1-x)^{\Lambda-\alpha_{q'}} 
 e^{-\frac{p_{t}}{\gamma_{q'}}}\!,
\label{fragm.eq}
 \end{eqnarray}
using Eqs.\ (\ref{alpha-u.eq}-\ref{alpha-us.eq}) for the relations between the
parameters  $\alpha_q$ and  the intercepts of the corresponding Regge
 trajectories. Here the large $x$ limit is defined by the probability to slow
 down (move away in rapidity space)
  the antiparton [(anti)quark or (anti)diquark] $\bar q'$ 
of the vacuum-created pair, as shown in  Fig.\ \ref{fig:fragm}~(left),
 while the small $x$ limit is obtained by slowing down both the original
 parton $q$ and the parton $q'$ attached to it to form the final hadron
 ($q+  q'\rightarrow h$), see Fig.\ \ref{fig:fragm}~(right) \cite{kai87}. 
 Various parton-antiparton pairs $\bar q'q'$
 created from the vacuum are sampled according to the corresponding probabilities
  $a_{q'}$, relative to $\bar u u$ and $\bar dd$ pairs ($a_u=a_d=1/2$).
 The iterative string fragmentation is continued until
  the mass  of the string
 reminder falls below a specified threshold\footnote{%More specifically, 
 We use $M_{\rm thr}^{qq'}=\sum _{i=1}^3 \sqrt{m_{h_i}^2+p^2_{t,{\rm thr}}}$,
where $h_1$,  $h_2$, and $h_3$ correspond to the minimal mass 3-hadron final
state  for the types $q$ and $q'$ of the string end  partons.}
 $M_{\rm thr}$,  followed by a modeling of
  a two particle decay of the remaining string.
  
The above-discussed treatment takes into account the production of final
``stable'' hadrons: pions, kaons, nucleons, lambda and sigma baryons,
including the corresponding antiparticles, where relevant. 
Contributions of decays of short-lived resonances are assumed here to be
accounted for in the stable hadron yields via the duality principle. The
exceptions are the formation of $\Delta^{++}$ and  $\Delta^{-}$ resonances
(and the corresponding antiparticles)
by a fragmentation of constituent $uu$ and $dd$ (anti)diquarks as well as the
production of  $\rho$ and $\eta$ mesons, which are also treated explicitly;
with the abundance of  $\rho$ mesons, relative to pions, and of  $\eta$,
relative to $\pi^0$, being controlled
by parameters  $w_{\rho}$ and $w_{\eta}$, respectively.
 
In case, a low mass  diffractive process is sampled, instead of an absorptive
interaction,  the mass of a 
diffractive state $M_X$ is defined according to the relativistic Breit-Wigner
distribution for the corresponding lowest mass resonance ($N^*$ for nucleons,
$\pi^*$ for pions, and $K_1$ for kaons), being matched at half-width to the
$dM_X/M_X^2$ dependence characteristic for the Pomeron-Pomeron-Reggeon  
($\mathbb{PPR}$) contribution.
The corresponding diffractive final state will consist, 
respectively,   of the products of the resonance decay or of hadrons
produced by a fragmentation of a short string, with valence partons
of the hadron (e.g., a valence quark and  a diquark, for the  proton case)
``sitting'' at the ends of this string. 
The transverse momentum transfer between the
projectile and the target states is modeled as
$\propto dp_t^2 \,e^{-p_t^2/\gamma_{\rm diff}}$, 
using a fixed\footnote{In principle, the shrinkage
of the diffractive cone should lead to some narrowing of the corresponding
distribution, i.e., to a decrease of $\gamma_{\rm diff}$ with energy.
Such an effect is neglected here for simplicity.} 
 parameter $\gamma_{\rm diff}$.

The generalization of the treatment  discussed in this Section to the case of hadron-nucleus
and nucleus-nucleus collisions is rather straightforward. The only difference to the hadron-proton case
 is that one considers multi-Pomeron exchanges between the projectile hadron
(nucleon) and any of the target nucleons. Correspondingly,
multiple strings of color field can be stretched between constituent partons
of a given hadron (nucleon) and the ones of a number of different nucleons of the partner
nucleus.

\subsection{Additional improvements  \label{prod-add.sec}}
In addition to the  basic procedure discussed in Section \ref{prod-basic.sec},
 we have a number of modifications of the treatment
of absorptive interactions, aimed at improving the description of
secondary hadron production in the projectile and target fragmentation regions.

First of all, this concerns the treatment of high mass diffraction (HMD), performed by
considering any of the cut Pomerons to be replaced by a diffractive cut of the 
triple-Pomeron graph corresponding to an interaction between three  soft Pomerons, with the  probability, for the particular case of projectile hadron HMD,
\begin{equation}
 w^{\rm HMD(proj)}_{hp(ij)}(s,b)=\chi_{hp(ij)}^{\rm HMD(proj)}(s,b)/\chi_{hp(ij)}^{\mathbb P}(s,b)\,.
 \label{eq:w-hmd} 
\end{equation}
Here the eikonal $\chi_{hp(ij)}^{\rm HMD(proj)}$ is defined as
\begin{equation}
 \chi_{hp(ij)}^{\rm HMD(proj)}(s,b)=\frac{G_{\mathbb{PPP}}}{8\pi}\int ^{s_0/\xi}_{\xi /s}\!
 \frac{dx_{\mathbb{P}}}{x_{\mathbb{P}}}\int \!d^2b'\;
 \chi_{h(i)}^{\mathbb P}(x_{\mathbb{P}}s,b')
\left(\chi_{p(j)}^{\mathbb P}(s_0/x_{\mathbb{P}},|\vec b - \vec b'|)\right)^2,
\label{eq:chi-hmd} 
\end{equation}
where $G_{\mathbb{PPP}}$ is the triple-Pomeron coupling, $\xi$ is the  minimal mass 
squared for a Pomeron, and $\chi_{h(i)}^{\mathbb P}$ is
the eikonal for a Pomeron exchange between   GW state $|i\rangle$ of  hadron $h$ 
and the triple-Pomeron vertex.

Using
\begin{equation}
\chi_{h(i)}^{\mathbb P}(\hat s,b)=\frac{\gamma_{h(i)}\,(\hat s/s_0)^{\Delta_{\rm s}}}
{R_{h(i)}^2+\alpha_{\mathbb P}'\,\ln (\hat s/s_0)}\;
 \exp \!\left[- \frac{b^2/4}{R_{h(i)}^2+\alpha_{\mathbb P}'\,\ln (\hat s/s_0)}\right],
\label{eq:chi-pom-leg}
\end{equation}
we have
\begin{eqnarray}
 \chi_{hp(ij)}^{\rm HMD(proj)}(s,b)=\frac{G_{\mathbb{PPP}}}{4}
 \int ^{s_0/\xi}_{\xi /s}\!
 \frac{dx_{\mathbb{P}}}{x_{\mathbb{P}}^{1+\Delta_{\rm s}}}\;
 \frac{\gamma_{h(i)}\,\gamma_{p(j)}^2\,  (s/s_0)^{\Delta_{\rm s}}}
{(R_{p(j)}^2-\alpha_{\mathbb P}'\,\ln x_{\mathbb{P}})
(R_{h(i)}^2+\frac{1}{2}R_{p(j)}^2+\alpha_{\mathbb P}'\,\ln  (\sqrt{x_{\mathbb{P}}}s/s_0))}
&& \nonumber \\
\times \; 
\exp \!\left[- \frac{b^2/4}{R_{h(i)}^2+\frac{1}{2}R_{p(j)}^2+\alpha_{\mathbb P}'\,
\ln  (\sqrt{x_{\mathbb{P}}}s/s_0)}\right]. &&
\label{eq:chi-hmd-int} 
\end{eqnarray}
$\chi_{hp(ij)}^{\rm HMD(targ)}$ for target HMD is obtained interchanging the indexes
$h(i)$ and $p(j)$ in the rhs of Eq.\ (\ref{eq:chi-hmd-int}).

In the considered case of projectile diffraction, instead
of a pair of strings stretched between constituent partons of the projectile and the
target, covering the full available rapidity range for the collision, $[0,Y]$,
$Y=\ln (s/s_0)$, we have ones covering the rapidity interval $[Y+\ln x_{\mathbb{P}},Y]$,
 leaving a rapidity gap  $[0,Y+\ln x_{\mathbb{P}}]$ devoid of secondary hadrons.
It is noteworthy that we obtain this way a diffractive interaction
only in the case when the above-discussed rapidity gap is not covered by secondary
hadrons emerging from other inelastic rescattering processes (cut Pomerons).
 For this to happen,
we either should have a single cut Pomeron configuration or, less likely, when all the
other cut Pomerons are represented by the same kind of diffractive contribution.

\begin{figure*}[t]
\centering
\includegraphics[height=4cm,width=0.3\textwidth]{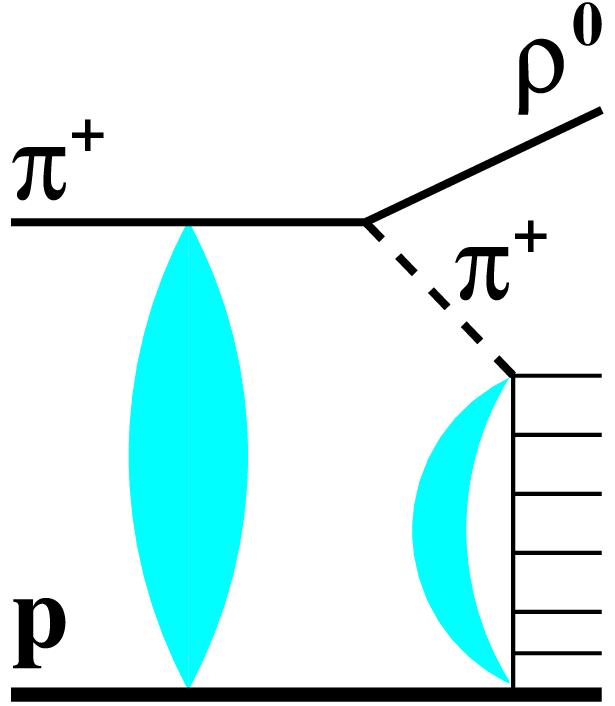}
\hspace{2cm}
\includegraphics[height=4cm,width=0.3\textwidth]{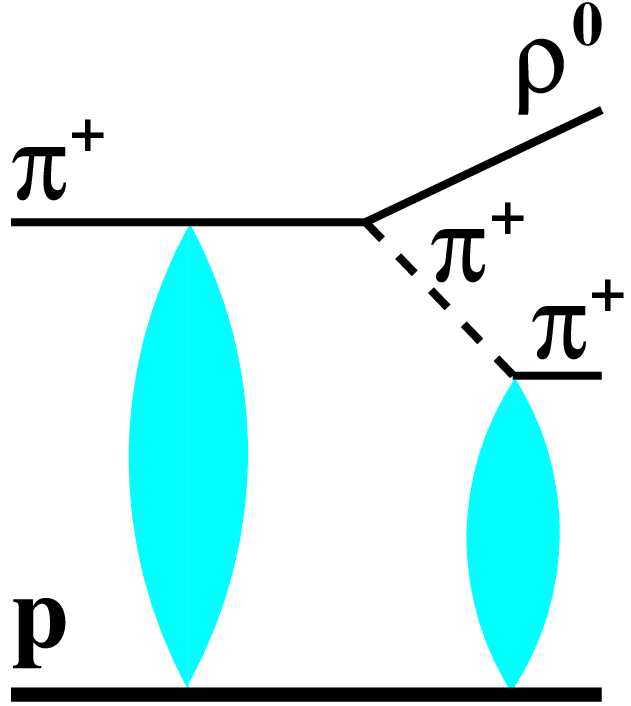}
\caption{Schematic view of the contributions of absorptive (left) and elastic (right) interactions of the virtual pion to $\pi^+p$ scattering.
The vertical ellipses correspond to the contributions of uncut Pomeron
exchanges. In particular, such exchanges between the incoming pion and 
the proton, shown by the large ellipses,
are responsible for absorptive corrections to the processes.}
\label{fig:piex}       
\end{figure*}

Further, in view of its importance for EAS muon number predictions,
  one has to take into account  $t$-channel pion exchange 
   in the Reggeon-Reggeon-Pomeron ($\mathbb{RRP}$) configuration \cite{ost13}.
   % (see, e.g., \cite{kai06}). 
   Here we improve the corresponding treatment of the  QGSJET-III model  \cite{ost21},  
 considering explicitly a rescattering of the virtual pion on the partner hadron (nucleus),
including the corresponding inelastic and elastic scattering contributions and 
accounting for   impact parameter dependent absorptive corrections for the process (see Fig.\  \ref{fig:piex}).
In such a case, we have a production of a leading hadron,
$h\stackrel{(\pi)}{\rightarrow}h'$, %$h+\pi \rightarrow h'$,
 and an interaction of the exchanged $t$-channel
virtual pion with the partner hadron (nucleus), where 
 $h'=p,n,\Delta^{++}$ for incident proton, $h'=\rho^{\pm}, \rho^0$
for incident $\pi^{\pm}$, and  $h'=K^*$ ($\bar K^*$) for incident kaon.
 
 The pion exchange contribution  is sampled according to the cross section 
 $\sigma^{\pi-{\rm exch}}_{hp}$ which, for the particular case of virtual pion
 emission by the projectile hadron, is defined as
 \begin{equation}
\sigma^{\pi-{\rm exch}}_{hp}(s)=g_{\pi/h} \int \! d^2b \; \sum_{i,j}C^{(i)}_{h}C^{(j)}_{p}\, 
S_{hp}^{(ij)}(s,b)\int \!dx\,dt\sum_{h'}f^{(\pi)}_{h\rightarrow h'}(x,t)\,
\Gamma ^{\rm tot}_{\pi p(j)}(\hat s_{\pi p},b)
\, .
\label{eq:sig-piex} 
\end{equation}
Here $g_{\pi/h}$ is an adjustable parameter which  controls the rate of the process and $f^{(\pi)}_{h\rightarrow h'}$ corresponds to the  distribution of
hadron $h'$, with respect to the  LC momentum fraction $x$ and  the squared momentum transfer $t$, in case of an energy-independent interaction of the
virtual pion with the target.
  $\Gamma ^{\rm tot}_{\pi p(j)}$ is the total (elastic plus inelastic)
profile for virtual pion interaction with the target proton represented by its
GW state $|j\rangle$ and $S_{hp}^{(ij)}$ is the so-called rapidity gap survival factor obtained by considering any number of uncut Pomeron exchanges between
the incoming hadron and the target proton (see Fig.\  \ref{fig:piex}),
 represented by GW states $|i\rangle$ and  $|j\rangle$, respectively:
\begin{equation}
S_{hp}^{(ij)}(s,b)=e^{-2\chi_{hp(ij)}^{\mathbb P}(s,b)}\,.
\label{eq:rgs} 
\end{equation}

The interaction profile  $\Gamma ^{\rm tot}_{\pi p(j)}$ is described by
(multi)Pomeron exchanges between the virtual pion and the target proton,
neglecting   Reggeon contributions:
\begin{equation}
\Gamma ^{\rm tot}_{\pi p(j)}(\hat s_{\pi p},b)=2 \sum_{i}C^{(i)}_{\pi}\left[1-e^{-\chi_{\pi p(ij)}^{\mathbb P}(\hat s_{\pi p},b)}\right],
\label{eq:pion-prof} 
\end{equation}
where the c.m.\ energy squared $\hat s_{\pi p}$ for the pion-proton interaction is
\begin{equation}
\hat s_{\pi p}=(1-x)(s-(m^2_{h'}+p_t^2)/x)\,.
\label{eq:pion-cme} 
\end{equation}
Here  the transverse momentum $p_t$ of the  hadron $h'$ is related to $t$ as
\begin{equation}
-t=p_t^2/x+(1-x)\,(m_{h'}^2/x-m_h^2)\,,
\label{eq:pion-t} 
\end{equation}
$m_h$ and $m_{h'}$ being the masses of $h$ and $h'$, respectively.

The type of hadron $h'$ as well as its  $x$ and   $p_t$   are sampled 
according to the product 
$f^{(\pi)}_{h\rightarrow h'}\,
\Gamma ^{\rm tot}_{\pi p(j)}$ [cf.\ the integrand in the rhs
of Eq.\ (\ref{eq:sig-piex})],
 with\footnote{In principle, the $t$-distribution of Eq.\ (\ref{eq:pion-xt})
 should be modified by  absorptive effects caused by
  the factor $S_{hp}^{(ij)}(s,b)$ in  Eq.\ (\ref{eq:sig-piex}).
   Such a modification  is
 neglected here for simplicity.} (e.g.\ \cite{kai06}):
\begin{equation}
f^{(\pi)}_{h\rightarrow h'}(x,t)= N^{-1}\, \frac{-t}{(t-m^2_{\pi})^2}\,
 (1-x)^{1-2\alpha_{\pi}(t)}\, e^{b_{\pi}t} .
\label{eq:pion-xt} 
\end{equation}
Here $N$ is the normalization constant ($\sum_{h'}\int \!dx\,dt\,f^{(\pi)}_{h\rightarrow h'}(x,t)=1$),
 $\alpha_{\pi}(t)=\alpha_{\mathbb R}'(t-m^2_{\pi})$ is the pion Regge  trajectory,
  % with the slope $\alpha_{\mathbb{R}}'\simeq 0.9$ GeV$^{-2}$,
  $m_{\pi}$ is the pion mass, and we use the same slope $b_{\pi}$ for virtual
  pion emission by any hadron.

With the probability 
$\Gamma ^{\rm el}_{\pi p(j)}(\hat s_{\pi p},b)/
\Gamma ^{\rm tot}_{\pi p(j)}(\hat s_{\pi p},b)$,
where\footnote{For simplicity, inelastic diffraction is neglected for virtual pion interactions.} 
\begin{equation}
\Gamma ^{\rm el}_{\pi p(j)}(\hat s_{\pi p},b)= \sum_{i}C^{(i)}_{\pi}\left[1-e^{-\chi_{\pi p(ij)}^{\mathbb P}(\hat s_{\pi p},b)}\right]^2,
\label{eq:pion-el-prof} 
\end{equation}
 we have an elastic scattering of the virtual pion, which simply
puts it on shell. Otherwise, we consider an absorptive interaction of the
 virtual pion, which  is modeled as described in  Section \ref{prod-basic.sec}.
 It is noteworthy that absorptive corrections push the pion exchange process
 towards large impact parameters, with increasing energy. This has two important consequences: i) the energy rise of $\sigma^{\pi-{\rm exch}}_{hp}$ is substantially slowed down,
  compared to the case when such absorptive 
 corrections are neglected (e.g., \cite{kai06}); ii) the probability of elastic scattering of the virtual pion
%  $\Gamma ^{\rm el}_{\pi p(j)}/\Gamma ^{\rm tot}_{\pi p(j)}$, 
  decreases with energy.

 Finally, regarding nondiffractive   production of various hadron species
 in the projectile and target fragmentation regions,
   we use a special treatment for the  first hadron emission off a string attached
to a ``remnant'' parton [diquark for a nucleon, light (anti)quark for a pion,
and strange (anti)quark for a kaon]. Namely,  we use the weights for a strange 
quark-antiquark pair, $a_{s/h}$, or for a diquark-antidiquark pair, $a_{ud/h}$,
 where relevant, creation from the vacuum, which differ from the standard
 string fragmentation parameters $a_s$ and $a_{ud}$, respectively.
  Phenomenologically, those can be interpreted
 as being related to   ``intrinsic strangeness'' in nucleons or in pions and
    ``intrinsic baryonic content'' of pions or kaons, correspondingly. While a more theoretically consistent approach would be to consider relevant Reggeon
 exchanges in the $\mathbb{RRP}$ configuration,
  with $\mathbb{R}=N, \Delta, K, K^*$, etc., similarly to the above-described
 pion exchange treatment, we prefer to use the discussed
  phenomenological implementation since it is more economic parameter-wise and more
  intuitively understandable for a potential non-expert model user.
  Further, in case a single cut Pomeron is attached to a  (anti)nucleon,
 the  first hadron emission off a string attached
to the ``remnant''  (anti)diquark is modified, compared to Eqs.\ (\ref{ems.eq}),
(\ref{alpha-ud.eq}):
replacing the parameter $\alpha_{qq}$ ($qq=uu$, $ud$, $dd$, or the corresponding 
antidiquarks) by  $\alpha_{qq}+\delta$, using an adjustable parameter $\delta$.

\subsection{Extension to low energies \label{prod-low.sec}}

Moving now to hadron-proton interactions at low energies, with the probability
\begin{equation}
w^{\rm inel({\mathbb R})}_{hp}(s)  =\sigma ^{\rm inel({\mathbb R})}_{hp}(s)/\sigma ^{\rm inel}_{hp}(s)\,,
\end{equation}
 we consider a Reggeon exchange dominated process.
 Otherwise, particle production is modeled as discussed in Sections
 \ref{prod-basic.sec} and  \ref{prod-add.sec}.
 
  Here $\sigma ^{\rm inel({\mathbb R})}_{hp}$  is obtained neglecting
 Pomeron contributions in Eqs.\ (\ref{Eq: A(s,b)}-\ref{sigel-b}):
 \begin{equation}
\sigma ^{\rm inel({\mathbb R})}_{hp}(s)= \int \!d^2b
\left[  2\Gamma ^{\mathbb R}_{hp}(s,b)-\left(\Gamma ^{\mathbb R}_{hp}(s,b)\right)^2\right]
=  \gamma_{hp}^{\mathbb R}\,(s/s_0)^{\alpha_{\mathbb R}-1}
   -\frac{\left[\gamma_{hp}^{\mathbb R}\,(s/s_0)^{\alpha_{\mathbb R}-1}\right]^2}
   {16\pi\,\lambda _{hp}^{\mathbb R}}\,.
\label{sigin-reg}
\end{equation}

\begin{figure*}[t]
\centering
\includegraphics[height=4cm,width=0.35\textwidth]{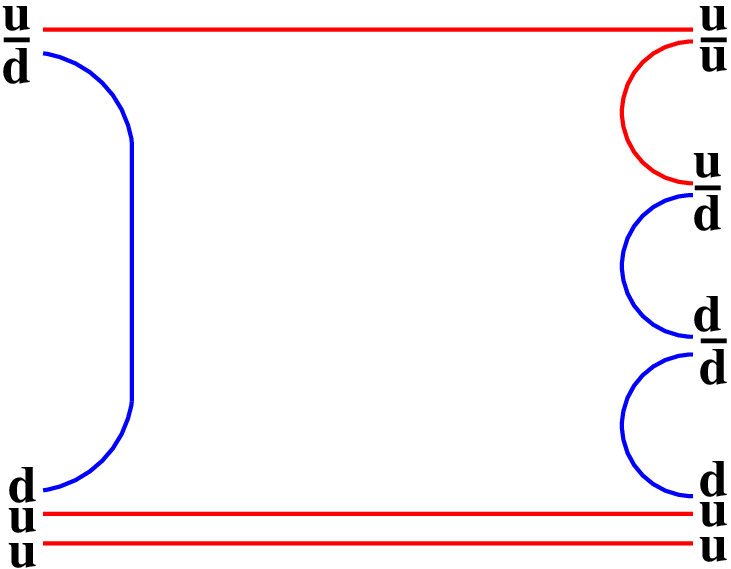}
\hspace{2cm}
\includegraphics[height=4cm,width=0.35\textwidth]{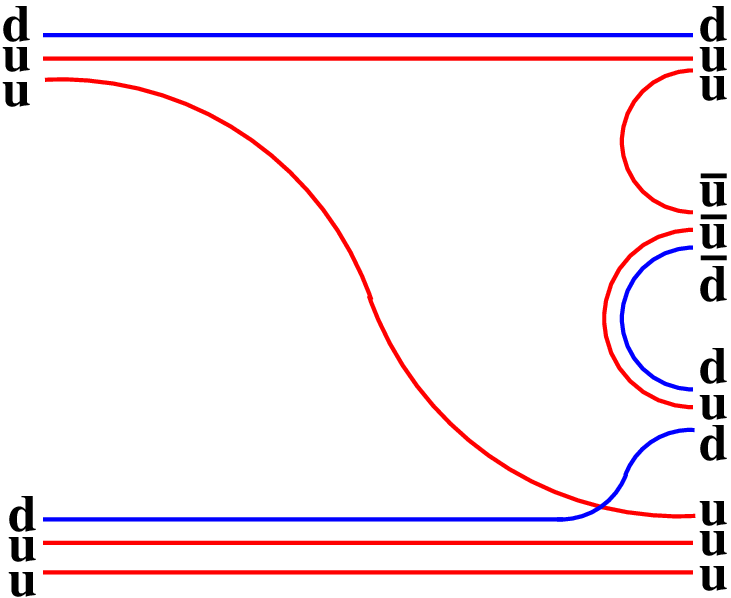}
\caption{Examples of a planar diagram contribution to hadron production in
$\pi^+p$ scattering (left) and of an undeveloped cylinder  diagram contribution to hadron production in $pp$ scattering (right).}
\label{fig:planar}       
\end{figure*}

For the Reggeon exchange dominated process, particle production is modeled as a 
fragmentation of a single string stretched between valence constituents of the
interacting hadrons. For $\bar pp$, $\pi ^{\pm} p$, and $K^-p$ scattering, 
 we consider the so-called planar diagram exemplified in Fig.\    \ref{fig:planar}~(left)
  for the case of $\pi ^+ p$ interaction: an antiquark of the incident hadron 
[$\bar d$  in Fig.\    \ref{fig:planar}~(left)]  annihilates with a valence quark of the target proton ($d$ quark in the Figure), such that a string of 
 color field is formed between the remaining valence parton of the projectile
 and the target diquark [$u$ and $uu$, respectively, in  Fig.\    \ref{fig:planar}~(left)]. On the other hand, since such a planar configuration 
 is  not possible for  $pp$ and $K^+p$ scattering, 
 we choose to describe  particle 
 production in these reactions
  by the so-called  ``undeveloped cylinder'' contribution exemplified
 in Fig.\    \ref{fig:planar}~(right) for the case of $pp$ interaction.
 In that case, a projectile valence quark [$u$ quark  in Fig.\    \ref{fig:planar}~(right)] is combined with a valence diquark
 ($uu$ in the Figure) to form a final baryon, while a string  of 
 color field is formed between the remaining valence parton of the projectile
 (a diquark for incident proton or $\bar s$ for $K^+$)
 and the target quark.\footnote{For $pp$ scattering, we consider also the 
 opposite configuration: when a target quark is combined with a projectile
 diquark and the string is stretched between a projectile quark and a target diquark.} The LC momentum partition between the (here, target) valence quark
 and the diquark is modeled like in Eq.\ (\ref{ems.eq}):
 $\propto x_{q_{\rm v}}^{-\alpha_{\mathbb{R}}}\, x_{qq}^{-\alpha_{qq}}\,
 \delta (1-x_{q_{\rm v}}-x_{qq})$.

For hadron-nucleus scattering, we define the probability
 $w^{\rm prod({\mathbb R})}_{hA}$ as corresponding to configurations of
 hadron-nucleus collisions, where all inelastic rescattering processes
 are described by Reggeon exchanges:
\begin{equation}
w^{\rm prod({\mathbb R})}_{hA}(s)  =\sigma ^{\rm prod({\mathbb R})}_{hA}(s)/\sigma ^{\rm prod}_{hA}(s)\,,
\end{equation}
where
\begin{eqnarray}
\sigma ^{\rm prod({\mathbb R})}_{hA}(s)
&=&  \int \! d^2b \;\sum_{i}C^{(i)}_{h}
\left\{\left[1-\sum_{j}C^{(j)}_{p}\int \!d^2\kappa \;T_A(\kappa) 
\right.\right. \nonumber \\
& \times & \left.  \left(1- e^{-2\chi^{\mathbb P}_{hp(ij)}(s,|\vec b-\vec \kappa |)}
- 2  \Gamma ^{\mathbb R}_{hp}(s,|\vec b-\vec \kappa |)
(1- e^{-\chi^{\mathbb P}_{hp(ij)}(s,|\vec b-\vec \kappa |)} )\right)\right]^A \nonumber \\
& -&\left[1-\sum_{j}C^{(j)}_{p}\int \!d^2\kappa \;T_A(\kappa)
 \left(1- e^{-2\chi^{\mathbb P}_{hp(ij)}(s,|\vec b-\vec \kappa |)} 
 + \Gamma ^{\mathbb R}_{hp}(s,|\vec b-\vec \kappa |)
\right.\right. \nonumber \\
& \times & \left.\left.\left.  
\left(2 e^{-\chi^{\mathbb P}_{hp(ij)}(s,|\vec b-\vec \kappa |)} 
-\Gamma ^{\mathbb R}_{hp}(s,|\vec b-\vec \kappa |)\right)\right)\right]^A
\right\},
\label{sigprod-ha-reg}
\end{eqnarray}
with $\Gamma ^{\mathbb R}_{hp}$ being defined in Eq.\ (\ref{eq:prof-rej}).
The corresponding  Reggeon exchange  dominated particle production is treated in a simplified fashion: as a single hadron-nucleon interaction, 
i.e., similarly to the hadron-proton case described above.

\section{Selected results \label{results.sec}}
\subsection{Hadron-proton and hadron-nucleus cross sections \label{results-cross.sec}}

Using various experimental data on hadron-proton interaction cross sections,
we fixed the model parameters describing Pomeron and Reggeon exchanges,
considering  $N_{\rm GW}=11$ Fock states per hadron,
 for the GW decomposition of hadron wave functions.\footnote{The model results
 depend rather weakly on the exact number of GW states considered.}
The obtained values are compiled in Tables \ref{Flo:param} and 
 \ref{Flo:param-reg}.
  \begin{table*}[t]
\begin{centering}
\begin{tabular}{|lllllllllll|}
\hline 
   $\Delta _{\rm s}$ &    $\Delta _{\rm sh}$ &   $\alpha'_{\mathbb{P}}$ & 
    $\delta _{\rm s/sh}$ &   $g_0$ &   $R^2_{p(1)}$ & $R^2_{\pi (1)}$ &
   $R^2_{K (1)}$    &  $d_p$ &  $d_{\pi}$ &   $d_{K}$
  \tabularnewline
&   &   {\scriptsize GeV$^{-2}$} &   &    {\scriptsize  GeV} &   {\scriptsize  GeV$^{-2}$} &   {\scriptsize  GeV$^{-2}$} &  {\scriptsize GeV$^{-2}$}  & & & 
  \tabularnewline
\hline 
0.06  &   0.22 &   0.2 & 0.92 & 1.63 & 2.81 &   1.32  &  1.11  & 0.1 & 0.17 &  0.17 
\tabularnewline
\hline
\end{tabular}\caption{Model parameters related to Pomeron exchange contributions.}
\label{Flo:param}
\par\end{centering}
 \end{table*}
 \begin{table*}[t]
\begin{centering}
\begin{tabular}{|llllllllllllll|}
\hline 
   $\alpha_{\mathbb R}$ &   $\alpha_{\mathbb R}'$ &   $\gamma^{\mathbb R}_{pp}$ & 
   $\gamma^{\mathbb R}_{\bar pp}$ &   $\gamma^{\mathbb R}_{\pi^+p}$ &   $\gamma^{\mathbb R}_{\pi^-p}$ &  $\gamma^{\mathbb R}_{K^+p}$ &
   $\gamma^{\mathbb R}_{K^-p}$    &  $\lambda^{\mathbb R}_{p}$ & $\lambda^{\mathbb R}_{\bar p}$  &    $\lambda^{\mathbb R}_{\pi^+}$ &    $\lambda^{\mathbb R}_{\pi^-}$
 &    $\lambda^{\mathbb R}_{K^+}$ &    $\lambda^{\mathbb R}_{K^-}$ 
  \tabularnewline
&    {\scriptsize GeV$^{-2}$}  &   {\scriptsize  GeV$^{-2}$} &  {\scriptsize  GeV$^{-2}$} &   {\scriptsize  GeV$^{-2}$}  &    {\scriptsize GeV$^{-2}$} &   {\scriptsize GeV$^{-2}$} &    {\scriptsize GeV$^{-2}$} &   {\scriptsize GeV$^{-2}$} & {\scriptsize GeV$^{-2}$}  &  {\scriptsize GeV$^{-2}$}  &  {\scriptsize GeV$^{-2}$}
 &  {\scriptsize GeV$^{-2}$}  &  {\scriptsize GeV$^{-2}$} 
  \tabularnewline
\hline 
0.5  &   0.9 &   5.5 & 11.5 & 3 & 4.1 &   0.8  &  3  & 1 & 3 &  0 &  0.4  &  0 &  0.4
\tabularnewline
\hline
\end{tabular}\caption{Model parameters related to Reggeon exchange contributions.}
\label{Flo:param-reg}
\par\end{centering}
 \end{table*}

\begin{figure*}[t]
\centering
\includegraphics[height=8cm,width=\textwidth]{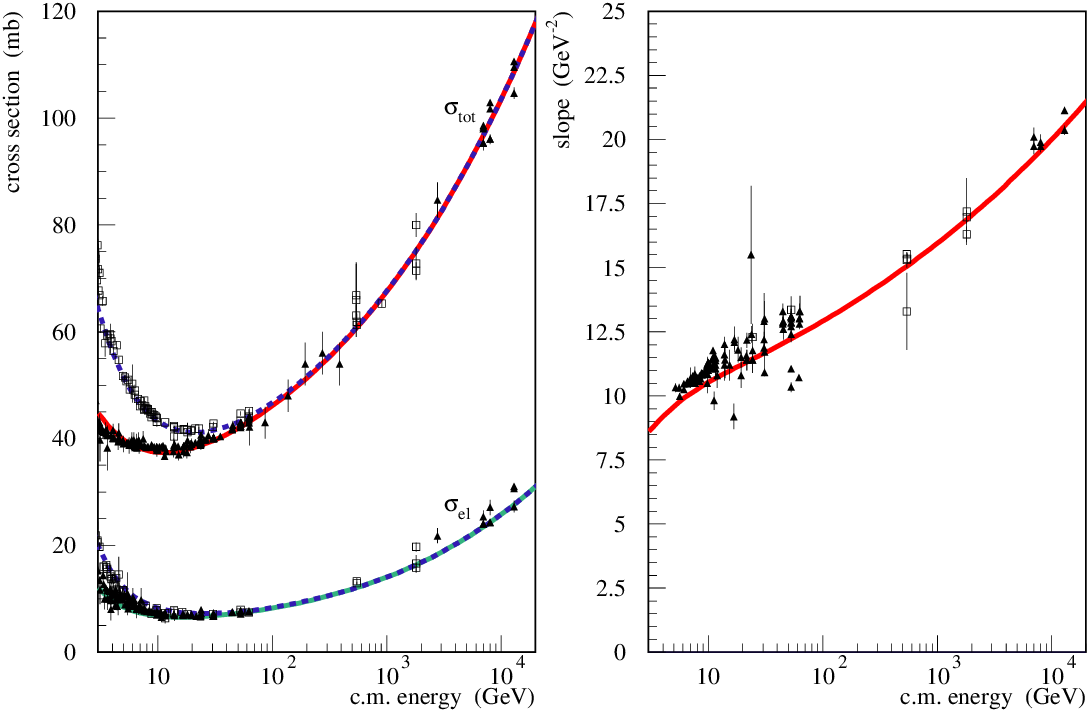}
\caption{C.m.\ energy dependence of   total and elastic 
cross sections for $pp$ and $\bar pp$ scattering  (left), and of 
  forward elastic  $pp$ scattering  slope (right), 
compared to experimental data \cite{pdg,ant19,aad23} (points): $pp$ -- red solid lines and filled triangles, $\bar pp$ -- blue dashed lines and open squares.}
 %from Refs.\ \cite{pdg,ant19,aad23}.}
\label{fig:sig-pp}       % Give a unique label
\end{figure*}

\begin{figure*}[t]
\centering
\includegraphics[height=8cm,width=\textwidth]{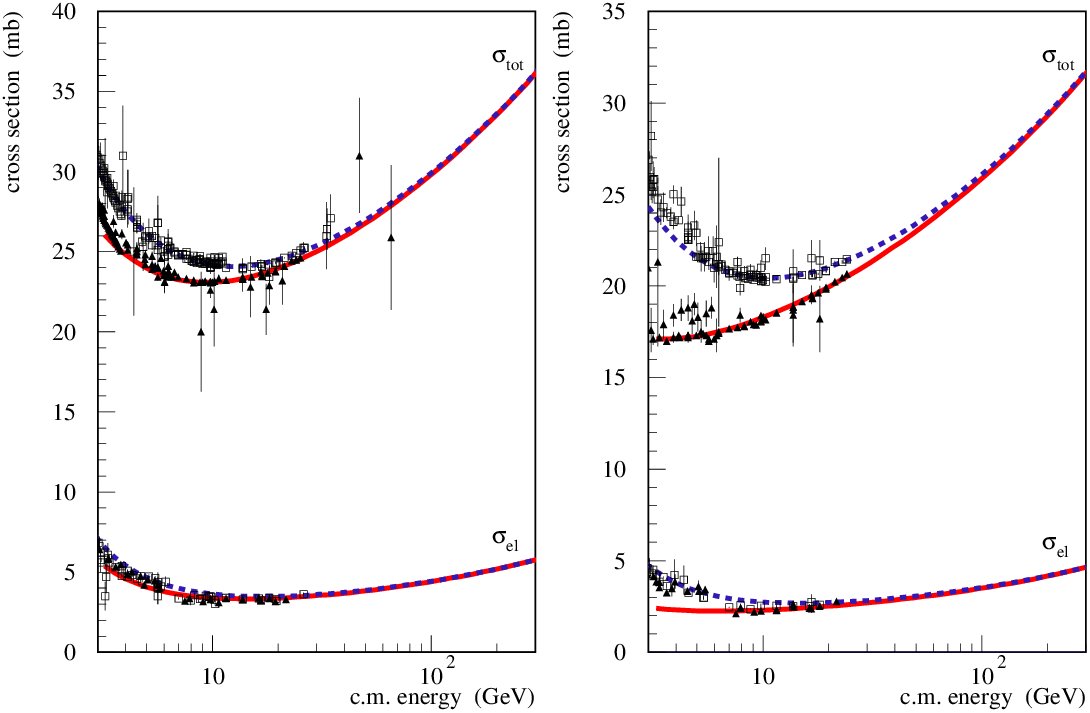}
\caption{C.m.\ energy dependence of  total and elastic 
cross sections for pion-proton   (left) and kaon-proton  (right) scattering, 
compared to experimental data \cite{pdg} (points): $\pi^+p$ and  $K^+p$ -- red solid lines and filled triangles; $\pi^-p$ and  $K^-p$ -- blue dashed lines and open squares.}
\label{fig:sig-pip}       % Give a unique label
\end{figure*}

 The corresponding  energy dependence of   total  and elastic proton-proton
and antiproton-proton cross sections  is compared
to experimental measurements in Fig.\ \ref{fig:sig-pp} (left), 
 while the one for   forward elastic proton-proton scattering slope,
%  $B^{\rm el}_{pp}=\left. d\ln (d\sigma^{\rm el}_{pp}/dt)/dt\right|_{t=0}$,
 $B^{\rm el}_{pp}$,   is plotted  in Fig.\ \ref{fig:sig-pp} (right).
% , in comparison to the corresponding accelerator  data.
 The correct description of the latter is also of importance since  $B^{\rm el}_{pp}$ is proportional to the average impact parameter squared for  $pp$ collisions.
 Thus, it   characterizes the ``fatness'' of the proton, i.e., the 
  transverse area occupied by its parton ``cloud'', which, in turn, impacts the multiple scattering rate in proton-proton and, especially,
proton-nucleus (nucleus-nucleus) collisions (see, e.g., the   discussion in \cite{ost19}).
Further in  Fig.\ \ref{fig:sig-pip}, the calculated energy dependence of  total  and elastic cross sections 
for pion-proton and kaon-proton scattering is confronted to experimental data.
 Overall, we have a good description of the experimental data over the 
 wide energy range considered.
 
\begin{figure*}[t]
\centering
\includegraphics[height=6cm,width=\textwidth]{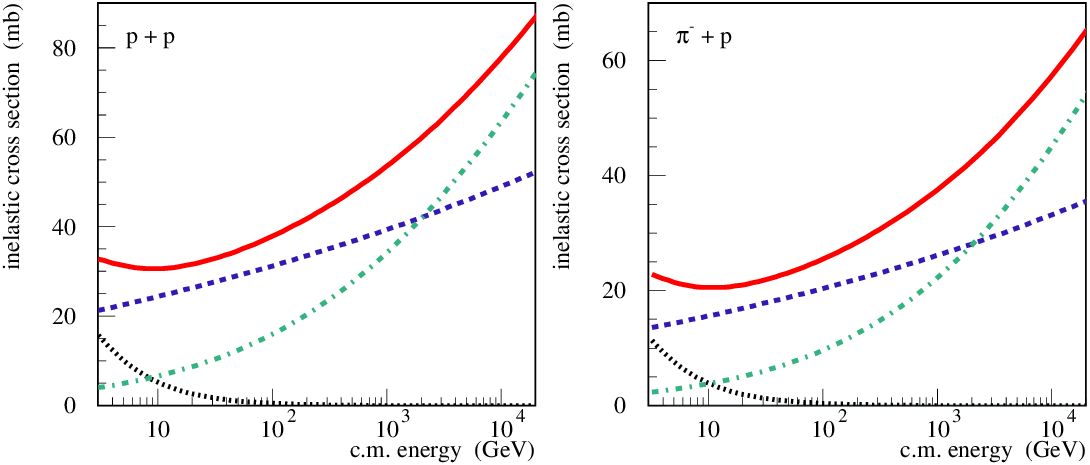}
\caption{C.m.\ energy dependence of   inelastic  $pp$ (left) and $\pi^-p$
(right) cross sections -- red solid lines.
 The results for $\sigma^{\rm inel}_{pp}$ and  $\sigma^{\rm inel}_{\pi^-p}$,
 obtained by taking into account only Reggeon, soft Pomeron, or semihard Pomeron contributions are plotted by black dotted, blue dashed, and green dash-dotted lines, respectively.}
\label{fig:sig-inel}       % Give a unique label
\end{figure*}

 In  Fig.\ \ref{fig:sig-inel}, we plot the energy dependence of 
 inelastic cross sections for $pp$ and $\pi ^-p$ collisions, showing also
  partial contributions of Reggeon-dominated interactions [i.e., setting
  $\chi^{\mathbb P}_{hp(ij)}\equiv 0$ in Eq.\ (\ref{Eq: A(s,b)})], as well as 
  $\sigma^{\rm inel}_{pp}$ and   $\sigma^{\rm inel}_{\pi ^-p}$
   calculated taking only soft or semihard Pomeron exchange into account. As one can see in the Figure, the Reggeon contribution
 steeply falls down with energy, becoming insignificant at $\sqrt{s}\gtrsim 20$
 GeV, while the semihard Pomeron 
 starts to dominate the inelastic cross sections at  $\sqrt{s}\gtrsim 2$ TeV.
 
\begin{figure*}[t]
\centering
\includegraphics[height=6cm,width=\textwidth]{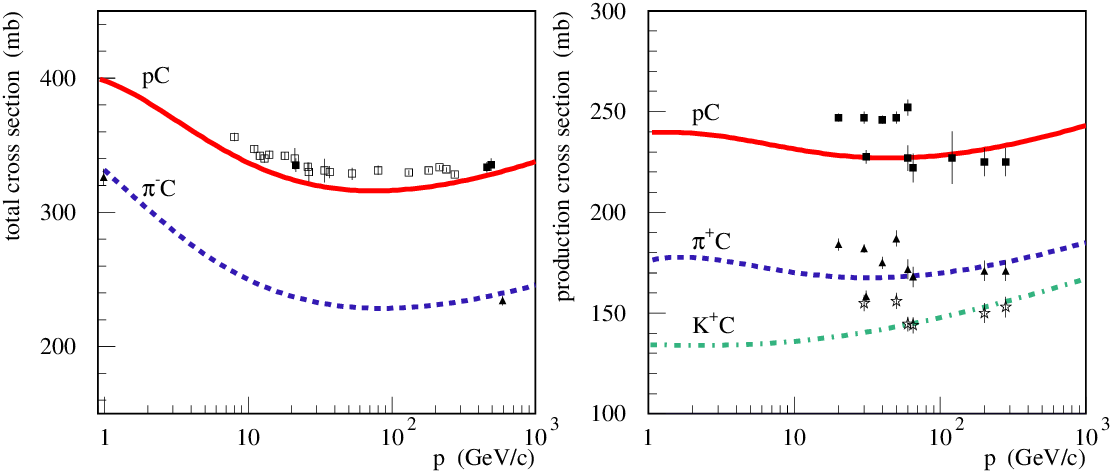}
\caption{Projectile lab.\ momentum dependence of   total cross sections for $p$C and   $\pi^-$C collisions (left) and of particle production  cross sections for $p$C,
 $\pi^+$C, and  $K^+$C interactions  (right), 
compared to experimental data
 \cite{bel66,eng70,jon71,den73,bab74,clo74,mur75,car79,der00,adu18,adu19,ach21}  (points):
$p$C -- red solid lines and filled squares, $\pi ^{\pm}$C  -- blue dashed lines and filled triangles,
 $K^{+}$C  -- green dash-dotted line and open stars; data on  $\sigma^{\rm tot}_{n{\rm C}}$ --
 open squares.}
\label{fig:sig-pa}       % Give a unique label
\end{figure*}

Further in  Fig.\ \ref{fig:sig-pa}, we present, as a function of lab.\ momentum
of incident hadron, the calculated 
 total cross sections  for proton-carbon and $\pi^-$C collisions,
 as well as  particle production cross sections
  for $p$C, $\pi^+$C, and $K^+$C  interactions.
   As we can see here, the Glauber-Gribov approach \cite{gri69,gla56} 
   is doing a very good job:
   using the model parameters tuned to hadron-proton cross section
   measurements, one obtains a good agreement with hadron-nucleus cross 
   section data.

\subsection{Secondary hadron production \label{results-prod.sec}}
As one can see  in  Fig.\ \ref{fig:sig-inel}, our treatment of hadronic interactions
 between $\sqrt{s}\simeq 20$ GeV and $\sqrt{s}\simeq 1$ TeV is dominated by soft
 Pomeron exchanges. This allowed us to fix the bulk of hadronization parameters
 of the model, compiled in Tables \ref{Flo:param-hadr} and \ref{Flo:param-sstring},  \begin{table*}[t]
\begin{centering}
\begin{tabular}{|llllllllllllll|}
\hline 
   $\alpha_{\rm sea}$ &   $\alpha_N(0)$ &    $\alpha_{\phi}(0)$ & 
   $\gamma_{\perp}$ &   $\gamma_{\perp}^p$ &   $\gamma_{\perp}^{\pi}$ &   $\gamma_{\perp}^{K}$ & $G_{\mathbb{PPP}}$ & $\xi$ & $\gamma_{\rm diff}$ & 
  $b_{\pi}$  &  $g_{\pi /p}$ & $g_{\pi /\pi}$ &   $g_{\pi /K}$
  \tabularnewline
 & & & {\scriptsize GeV$^2$} & {\scriptsize GeV$^2$} & {\scriptsize GeV$^2$} & {\scriptsize GeV$^2$} & {\scriptsize GeV} & {\scriptsize GeV$^2$} & {\scriptsize GeV$^2$} &  {\scriptsize GeV$^{-2}$} & &  &
  \tabularnewline
\hline 
0.5 & -0.05 & 0 & 0.15 & 0.75 & 0.2 &  0.2  &  0.14 & 4.5 & 0.41 &  0.3 & 0.15 &
  0.15 &  0.05
\tabularnewline
\hline
\end{tabular}\caption{Parameters of the hadronization procedure (excluding string fragmentation parameters).}
\label{Flo:param-hadr}
\par\end{centering}
 \end{table*}
\begin{table*}[t]
\begin{centering}
\begin{tabular}{|lllllllllllllll|}
\hline 
 $\Lambda$ & $p^2_{t,{\rm thr}}$ & $a_{ud}$ & $a_{s}$ &  $a_{us}=a_{ds}$ & $a_{s/p}$ & $a_{s/\pi}$
 & $a_{ud/\pi}$  & $a_{ud/K}$ & $\delta$ & $w_{\eta}$ & $w_{\rho}$ & $\gamma_{qq}$ &    $\gamma_{u}=\gamma_{d}$  & $\gamma_{s}$
  \tabularnewline
 & {\scriptsize GeV$^2$} & & & & & & & & & & &  {\scriptsize GeV$^2$}  &  {\scriptsize GeV$^2$}  &  {\scriptsize GeV$^2$} 
  \tabularnewline
\hline 
1.7 & 0.3 & 0.1 & 0.1 & 0.019 & 0.21 & 0.19 & 0.19 & 0.1 & 1 & 0.12 & 0.4 & 0.27 & 0.18 & 0.24
\tabularnewline
\hline
\end{tabular}\caption{String fragmentation parameters for soft Pomerons.}
\label{Flo:param-sstring}
\par\end{centering}
 \end{table*}
 using experimental data on hadron production in proton-proton, pion-proton, and
 kaon-proton collisions in this energy range.

\begin{figure*}[t]
\centering
\includegraphics[height=12cm,width=\textwidth]{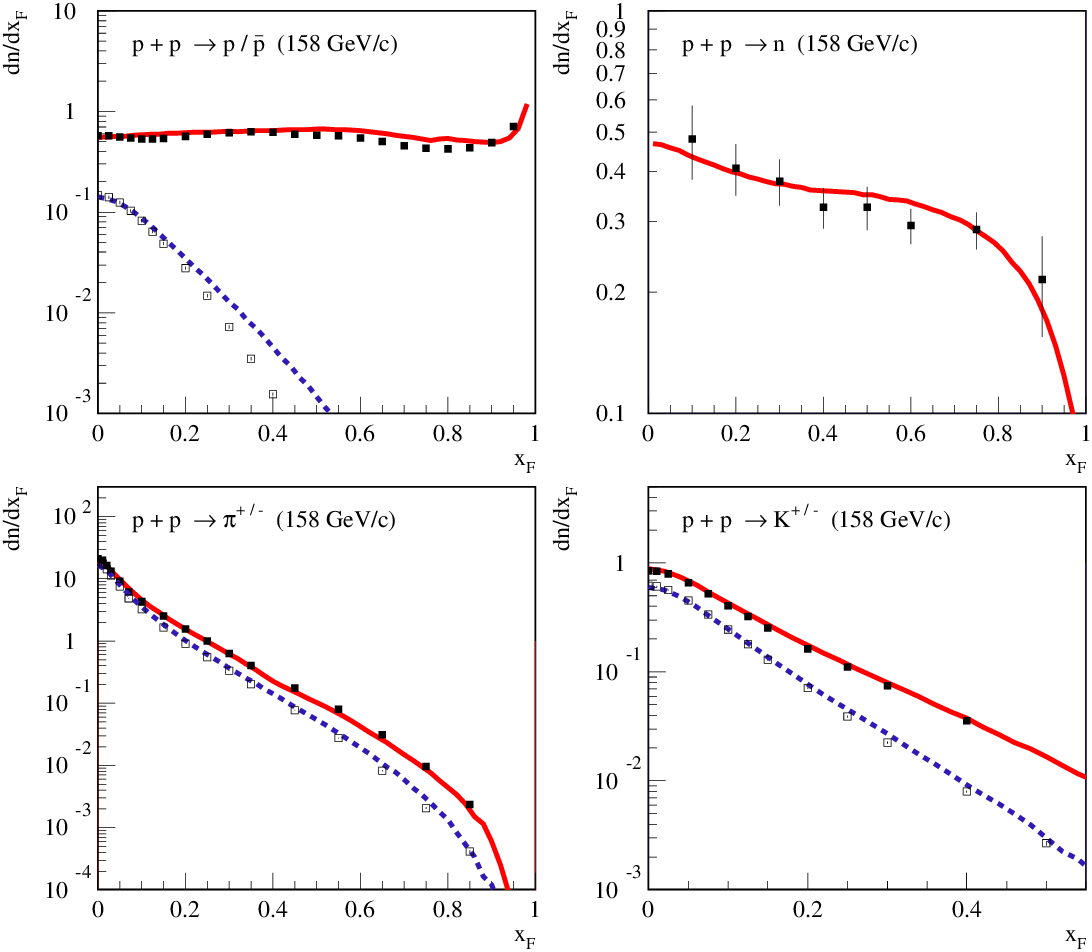}
\caption{Feynman $x$ distributions of protons and antiprotons (top left), neutrons
(top right), charged pions (bottom left), and charged kaons (bottom right),
produced in $pp$ collisions at 158 GeV/c, compared to NA49 data \cite{alt06,ant10,ant10a} (points): $p$, $n$, $\pi^+$, and  $K^+$ -- red solid lines and filled  squares;  $\bar p$, $\pi^-$ and  $K^-$ -- blue dashed lines and open squares.}
% The contribution of the pion exchange process to neutron production is plotted by green dash-dotted line.}
\label{fig:pp158}       % Give a unique label
\end{figure*}

In Fig.\ \ref{fig:pp158}, we compare our calculations of  
 Feynman $x$ distributions of secondary hadrons produced in proton-proton interactions, for 158 GeV/c lab.\ momentum of incident proton,  to the corresponding data of the NA49 experiment. As one can see in the Figure,
 we obtained a good description of the measurements, with the exception
 of the antiproton spectrum, which is substantially harder than the
  observed one.
 
\begin{figure*}[t]
\centering
\includegraphics[height=6cm,width=\textwidth]{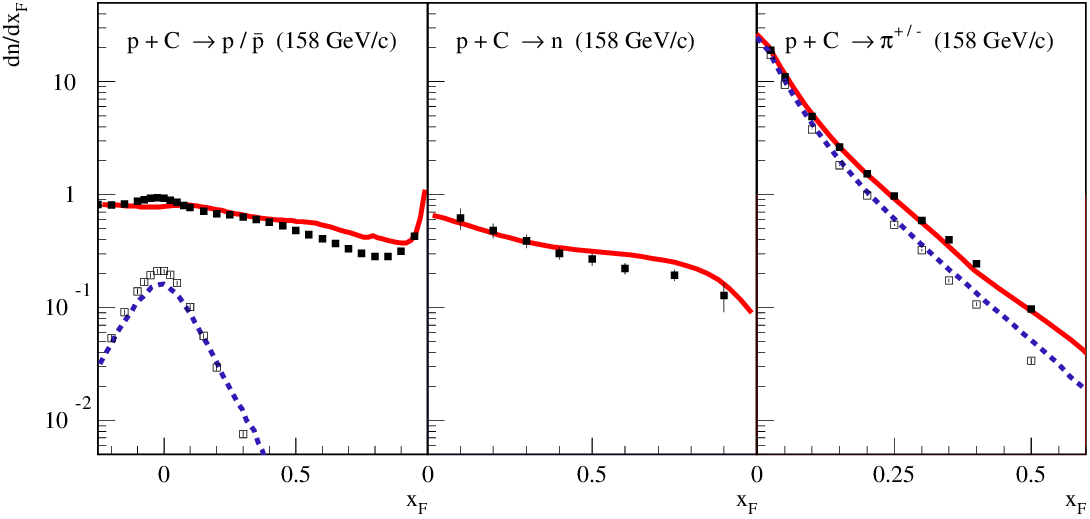}
\caption{Feynman $x$ distributions of protons and antiprotons (left panel), neutrons
(middle panel), and charged pions (right panel),
produced in $p$C collisions at 158 GeV/c, compared to NA49 data \cite{alt07,baa13} (points): $p$, $n$, and $\pi^+$ -- red solid lines and filled  squares; $\bar p$ and  $\pi^-$ 
-- blue dashed lines and open squares.}
% The contribution of the pion exchange process to neutron production  is plotted by green dash-dotted line.}
\label{fig:pc158}       % Give a unique label
\end{figure*}

 A similar comparison for  hadrons produced in proton-carbon collisions, also at 158 GeV/c,    is presented in  Fig.\ \ref{fig:pc158}.
 In contrast to the $pp$ case in  Fig.\ \ref{fig:pp158}, here we observe a certain
 excess of protons and neutrons produced at large values of  Feynman $x$,
 compared to  NA49 data.
 This may indicate that the model underestimates somewhat the inelasticity,
 i.e., the relative energy loss of   leading (most energetic) secondary
 nucleons, in proton-nucleus collisions. 

\begin{figure*}[t]
\centering
\includegraphics[height=19.cm,width=\textwidth]{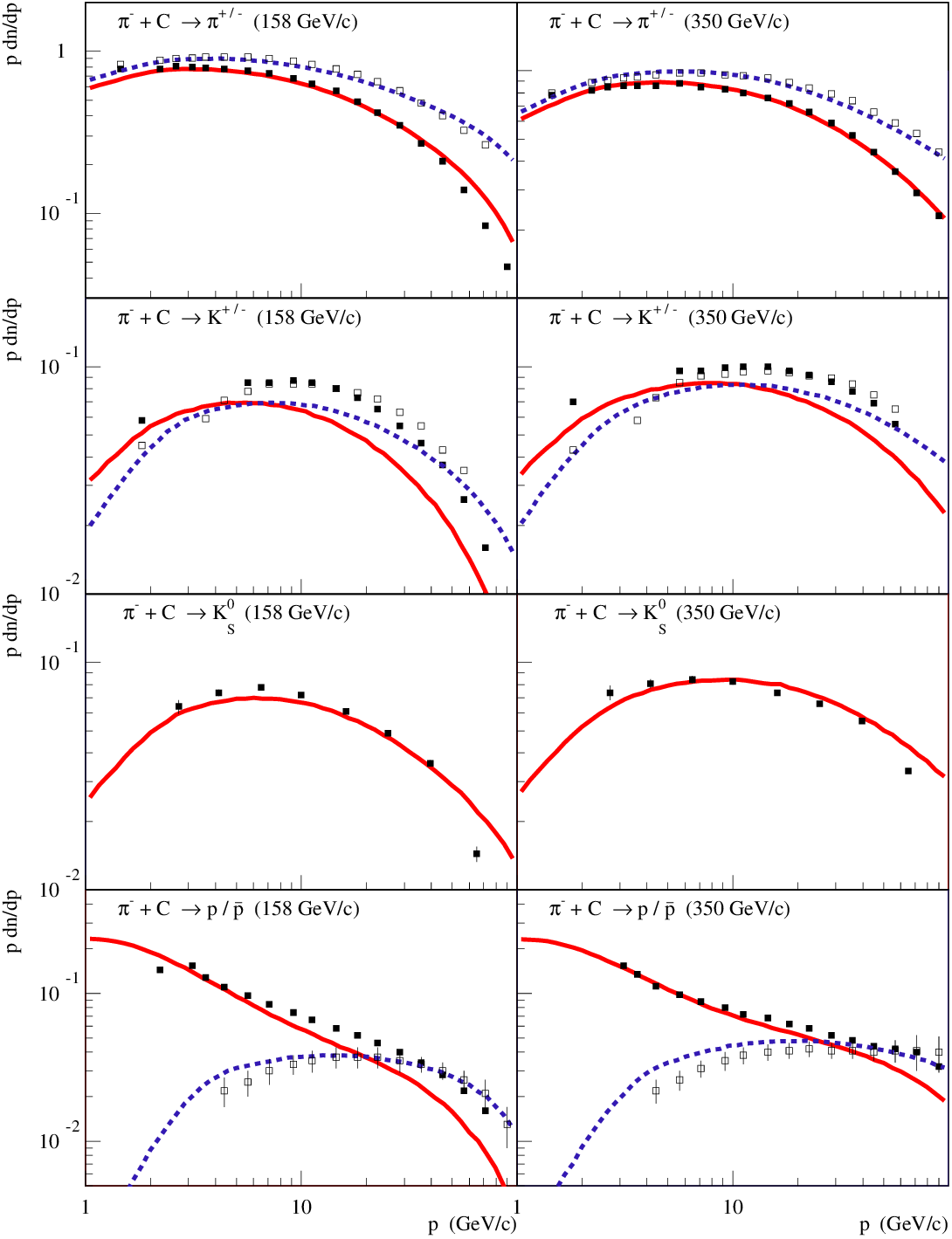}
\caption{Lab.\ momentum distributions of, from top to bottom, $\pi^{\pm}$,  $K^{\pm}$,  $K_{\rm S}^0$, $p$ and $\bar p$, produced in $\pi^-$C collisions at 158 GeV/c (1st column) and  350 GeV/c (2nd column),
 compared to NA61/SHINE data \cite{adh23} (points):
 $p$, $\pi^+$,  $K^+$, and  $K_{\rm S}^0$  -- red solid lines and filled  squares;
  $\pi^-$,  $K^-$, and   $\bar p$ -- blue dashed lines and open squares.}
\label{fig:pimc158}       % Give a unique label
\end{figure*}

In turn, in Fig.\ \ref{fig:pimc158}, the calculated lab.\  momentum distributions of
various hadrons produced in $\pi^-$C collisions at 158 GeV/c  and  350 GeV/c
incident pion momenta are compared to measurements by the NA61/SHINE experiment.
While our results agree with the observed spectra of charged pions
reasonably well, the situation with   kaon production is less clear.
On the one side, the calculated charged kaon yields are noticeably smaller
than the experimental ones. On the other hand, no such  ``kaon deficiency''
is observed for $K_{\rm S}^0$. The NA61/SHINE collaboration interpreted such an
enhancement of charged kaon production, relative to neutral kaon yields,
as an indication of a substantial violation of the isospin symmetry \cite{adh25}. 
However, such a radical assumption may not be necessary, considering
a general difficulty to discriminate $K^{\pm}$ from abundant charged pions,
compared to a more reliable particle identification for   $K_{\rm S}^0$.
Similar concerns apply to the measured spectra of protons and antiprotons.
Tuning the model to the corresponding  NA61/SHINE data, as plotted in 
the bottom row of Fig.\ \ref{fig:pimc158}, we arrive to a
 substantial overestimation of   $p$ and $\bar p$ yields, compared to 
results of other experiments (see, e.g., \cite{ost24d} for the corresponding discussion).

\begin{figure*}[t]
\centering
\includegraphics[height=6cm,width=\textwidth]{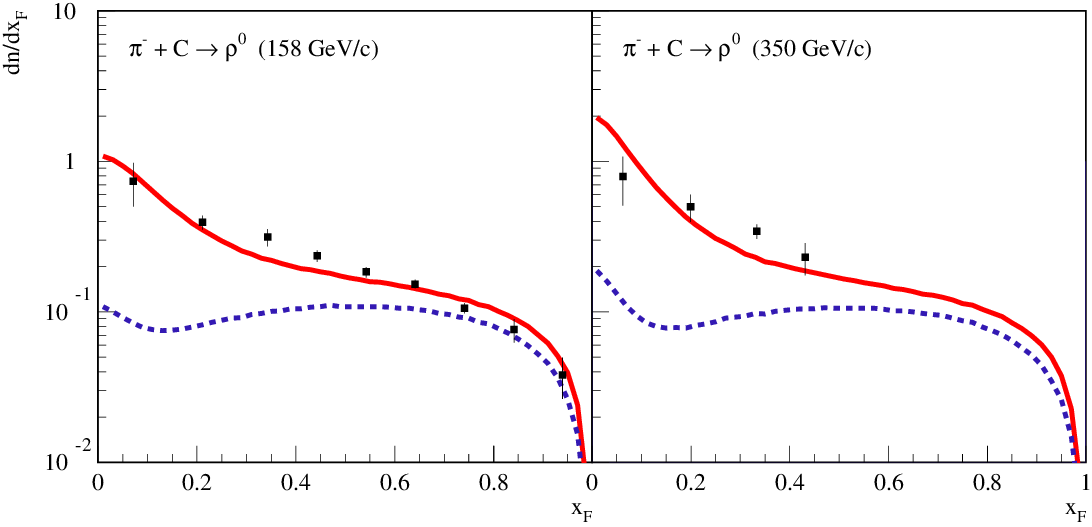}
\caption{Feynman $x$ distributions of $\rho^0$ mesons
produced in $\pi^-$C collisions at 158 GeV/c (left) and 350 GeV/c (right) --
red solid lines, compared to NA61/SHINE  data \cite{adu17} (points).
Shown by blue dashed lines are partial contributions of the pion exchange process, corresponding to virtual pion emission by the incident pion.}
\label{fig:pi-rho}       % Give a unique label
\end{figure*}

Of particular importance are measurements of forward $\rho^0$ meson production
 by the NA61/SHINE experiment \cite{adu17}, which allowed us to constrain the
  rate of the pion exchange 
 process in pion-proton and pion-nucleus collisions by tuning  the corresponding
 parameter  $g_{\pi /\pi}$. The calculated Feynman $x$ distributions of $\rho^0$ mesons,
for $\pi^-$C collisions at 158 and  350 GeV/c, are compared to the NA61/SHINE data
in Fig.\ \ref{fig:pi-rho}.
 As one can see in the Figure, the contribution of the pion exchange dominates forward
  $\rho^0$ meson production, having thereby a strong impact on predictions
   for   muon content of extensive air showers \cite{ost13}.
  
\begin{figure*}[t]
\centering
\includegraphics[height=11.cm,width=\textwidth]{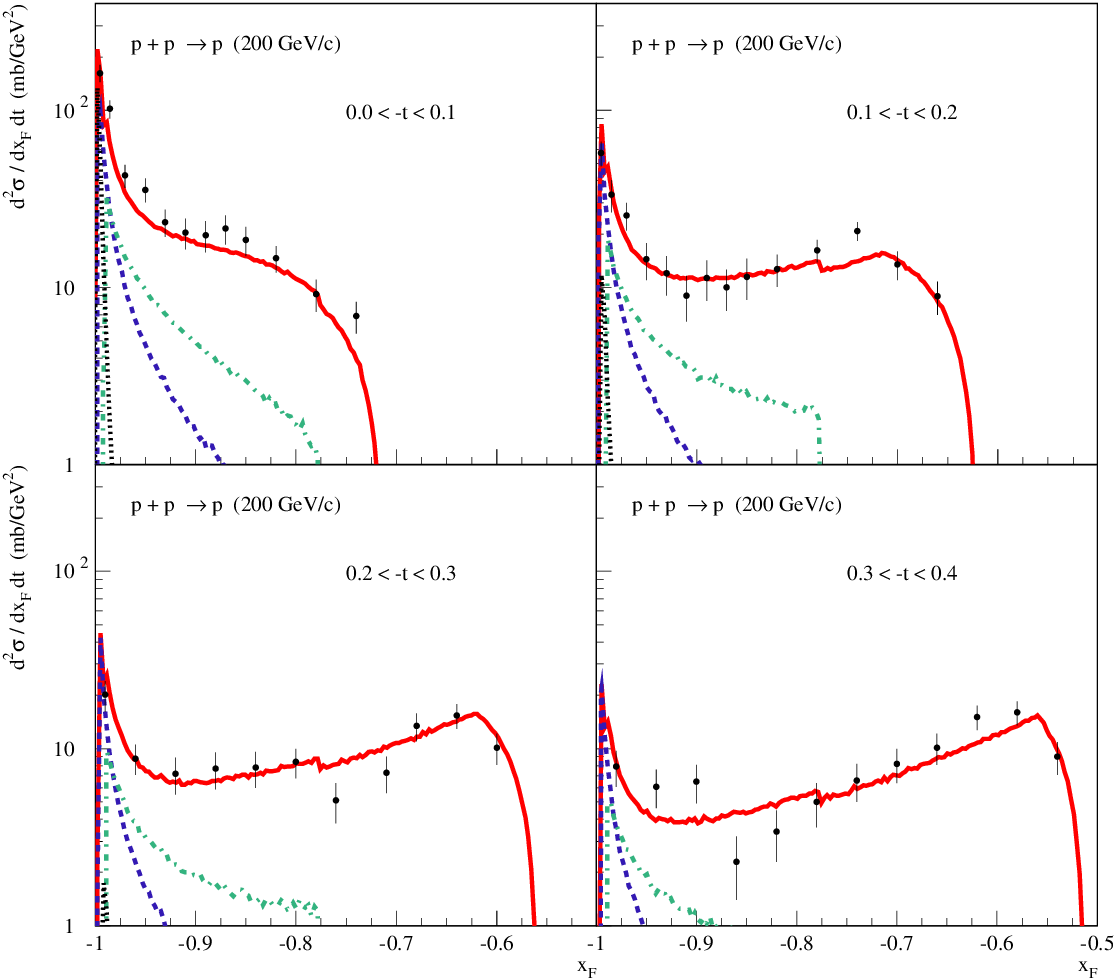}
\caption{Feynman $x$ dependence of double differential cross sections, 
$d^2\sigma /dx_{\rm F}/dt$,  for proton production in the target fragmentation
 region in $pp$   interactions at 205 GeV/c, for different ranges of $t$, as indicated
 in the plots  (red solid lines),
  compared to experimental  data \cite{whi75} (points).
Partial contributions of low and high mass diffraction of   incident proton
are shown by blue dashed and green dash-dotted lines, respectively, while 
the contributions  of the pion exchange process, corresponding to elastic scattering of the  virtual pion emitted by  incident proton, are plotted 
as black dotted lines.}
\label{fig:pdiffrt}       % Give a unique label
\end{figure*}

\begin{figure*}[t]
\centering
\includegraphics[height=6cm,width=\textwidth]{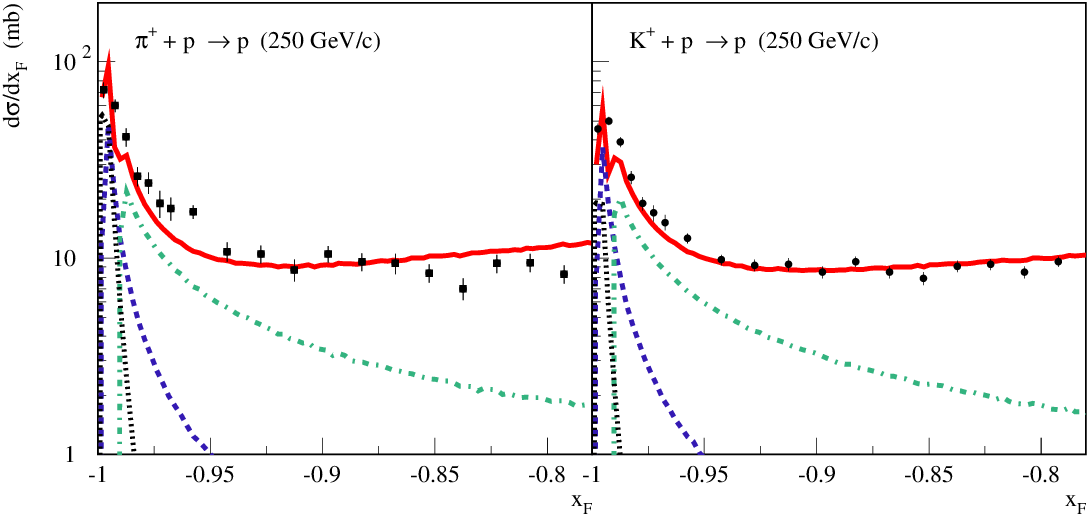}
\caption{Feynman $x$ dependence of inclusive cross sections  for proton production in the target fragmentation region in $\pi^+p$ (left) and  $K^+p$ (right) 
interactions at 250 GeV/c (red solid lines),
 compared to experimental  data \cite{aji89} (points).
Partial contributions of low and high mass diffraction of   incident hadrons
are shown by blue dashed and green dash-dotted lines, respectively, while 
the contributions  of the pion exchange process, corresponding to elastic scattering of the  virtual pion emitted by  incident hadrons, are plotted 
as black dotted lines.}
\label{fig:hdiffrt}       % Give a unique label
\end{figure*}

  Also of importance are measurements of secondary proton spectra in the target fragmentation region, in $pp$, $\pi^+p$, and $K^+p$ interactions \cite{whi75,aji89},
   which allowed us to constrain the rate of  low mass diffraction
  of  projectile hadrons. Our calculations are compared in  Fig.\ \ref{fig:pdiffrt}
    to the corresponding data for proton-proton collisions  at 205 GeV/c,
  for different selections of   momentum transfer
  squared $t$ for target proton. In turn, in Fig.\  \ref{fig:hdiffrt}, the calculated
  proton spectra in the target fragmentation region  in  $\pi^+p$ and $K^+p$ interactions at 250 GeV/c  are compared to the measured ones.  As one can see in the Figures, significant contributions to  backward proton production come both
 from low and high mass diffraction of incident hadrons. Moreover, the 
 spectral peak at $x_{\rm F}\simeq -1$, for small values of  $t$, is dominated by the
  contribution of the pion exchange process, corresponding to elastic scattering of
   the virtual pion  emitted by the  projectile.

Regarding high mass diffraction, the corresponding parameter, the triple-Pomeron
coupling $G_{\mathbb{PPP}}$, has been fixed based on measured rates of single
diffractive-like events in $pp$ collisions at  $\sqrt{s}=7$ TeV. The calculated
rates are compared  in Table \ref{tab: SD-totem} 
 to the corresponding data of the TOTEM experiment, for
different intervals of    mass $M_{X}$ of diffractive states produced.
\begin{table*}[t]
\begin{tabular*}{1\textwidth}{@{\extracolsep{\fill}}lllll}
\hline 
$M_{X}$ range, GeV & $3.1-7.7$  &  $7.7-380$  & $380-1150$  &  $1150-3100$ \tabularnewline
\hline 
\hline 
TOTEM   & $1.83\pm 0.35$ & $4.33 \pm 0.61$ & $2.10 \pm 0.49$
 & $2.84 \pm 0.40$  \tabularnewline
QGSb & 1.84 & 4.37 & 1.55 & 5.19 \tabularnewline
\hline 
\end{tabular*}
\caption{Cross sections of single diffractive-like  events (in mb),
for $pp$ interactions  at $\sqrt{s}=7$ TeV, for different ranges of mass $M_{X}$ of diffractive states produced,  compared to TOTEM data  \cite{olj20}.
\label{tab: SD-totem}}
\end{table*}
Obviously, the agreement between our results and the data is quite good for
all  $M_{X}$ ranges, except the last one corresponding to large
  $M_{X}>1.15$ TeV.\footnote{As discussed in \cite{ost24b}, the production of
   high mass diffractive-like states
is dominated by nondiffractive background.}

\begin{figure*}[t]
\centering
\includegraphics[height=19cm,width=\textwidth]{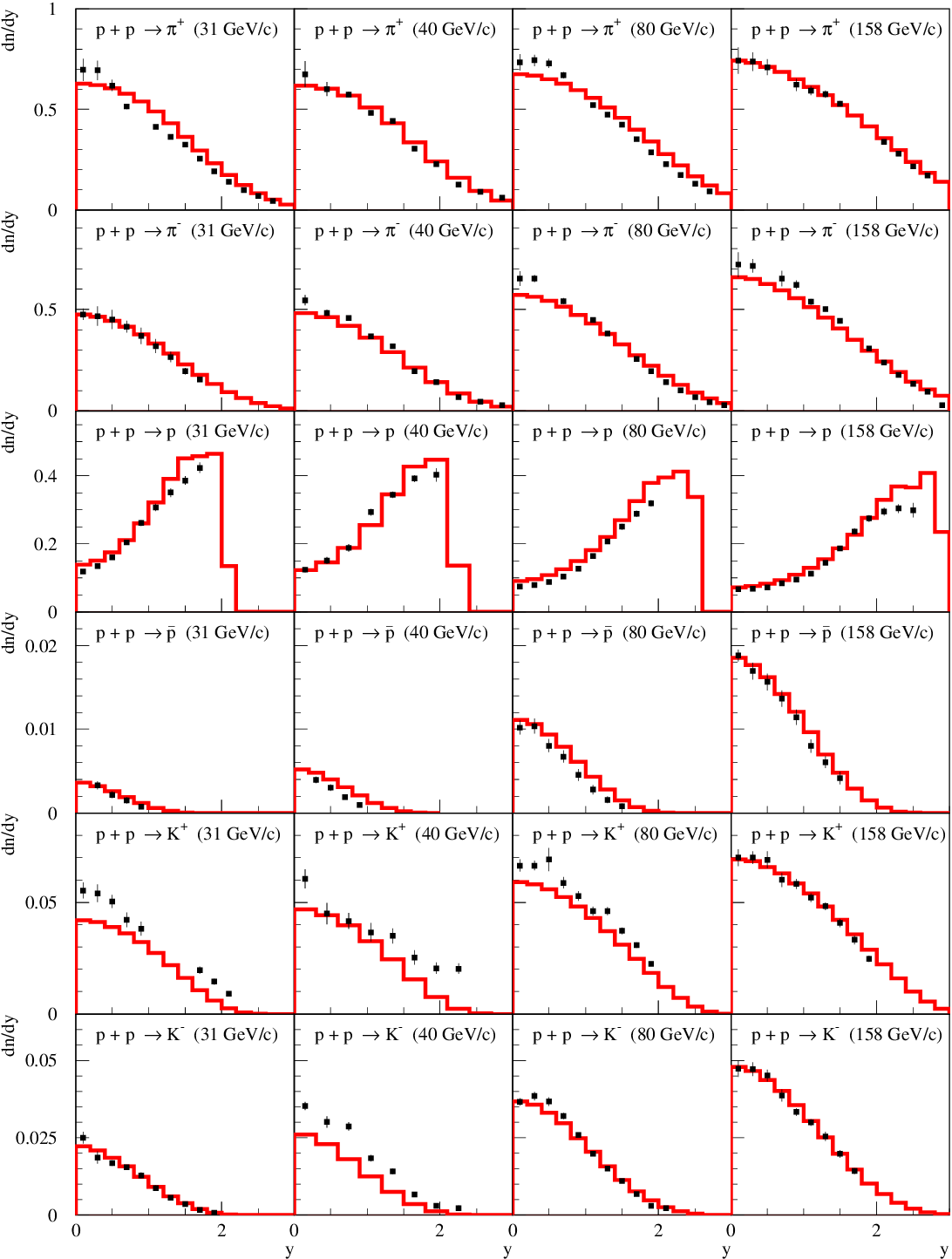}
\caption{Rapidity distributions (in c.m.\ frame) of, from top to bottom, $\pi^{+}$, $\pi^{-}$, $p$,
 $\bar p$,  $K^{+}$, and   $K^{-}$, produced in $pp$ collisions
  at 31, 40, 80, and 158 GeV/c, as indicated  in the plots, 
 compared to NA61/SHINE data \cite{adu17a} (points).}
\label{fig:pp-na61}       % Give a unique label
\end{figure*}

\begin{figure*}[t]
\centering
\includegraphics[height=19cm,width=\textwidth]{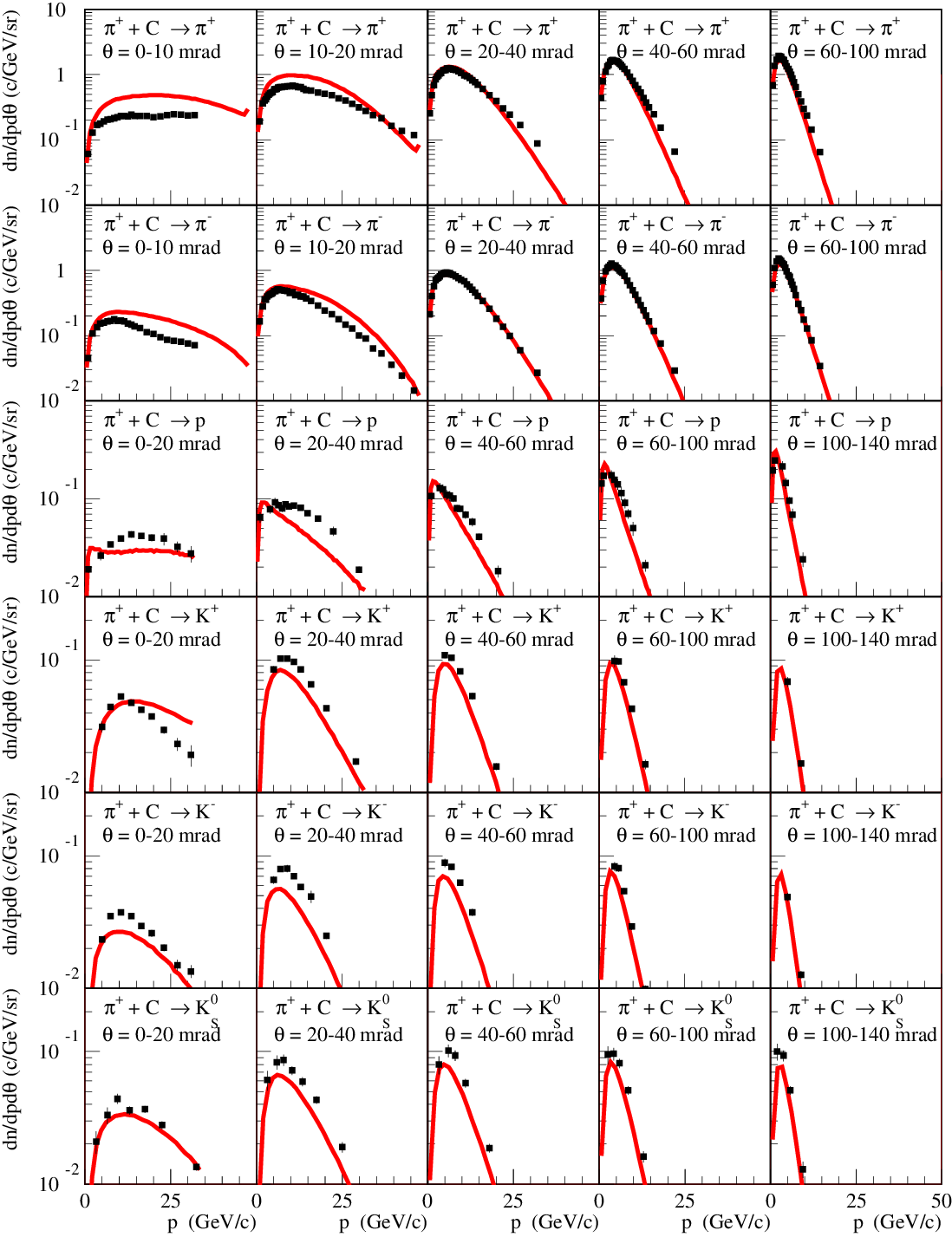}
\caption{Lab.\ momentum distributions of, from top to bottom, $\pi^{+}$, $\pi^{-}$, $p$,
  $K^{+}$,    $K^{-}$, and  $K_{\rm S}^0$, produced in $\pi^+$C collisions at 60 GeV/c, at different polar angles, as indicated in the plots,
 compared to NA61/SHINE data \cite{adu19a} (points).}
\label{fig:pipc60}       % Give a unique label
\end{figure*}

Moving now to lower energies, where the Reggeon exchange contribution becomes
important,  we have a reasonable agreement between the calculated
  rapidity distributions of $p$, $\bar p$,
 $\pi^{\pm}$, and  $K^{\pm}$, for $pp$ interactions at 31, 40, 80, and 158 GeV/c,
plotted in Fig.\ \ref{fig:pp-na61},
 and the corresponding results of the  NA61/SHINE experiment.
In turn, in Fig.\ \ref{fig:pipc60}, we compare our calculations for lab.\ momentum  %  
 distributions of secondary hadrons produced in $\pi^+$C collisions at 60 GeV/c,
 at different polar angles, to the respective
data  of   NA61/SHINE. While the overall agreement with the
measurements remains at an  acceptable level, we clearly have an excess of 
charged pion production at   small angles, $\theta \lesssim 20$ mrad,
compared to the data.

\begin{figure*}[t]
\centering
\includegraphics[height=19cm,width=\textwidth]{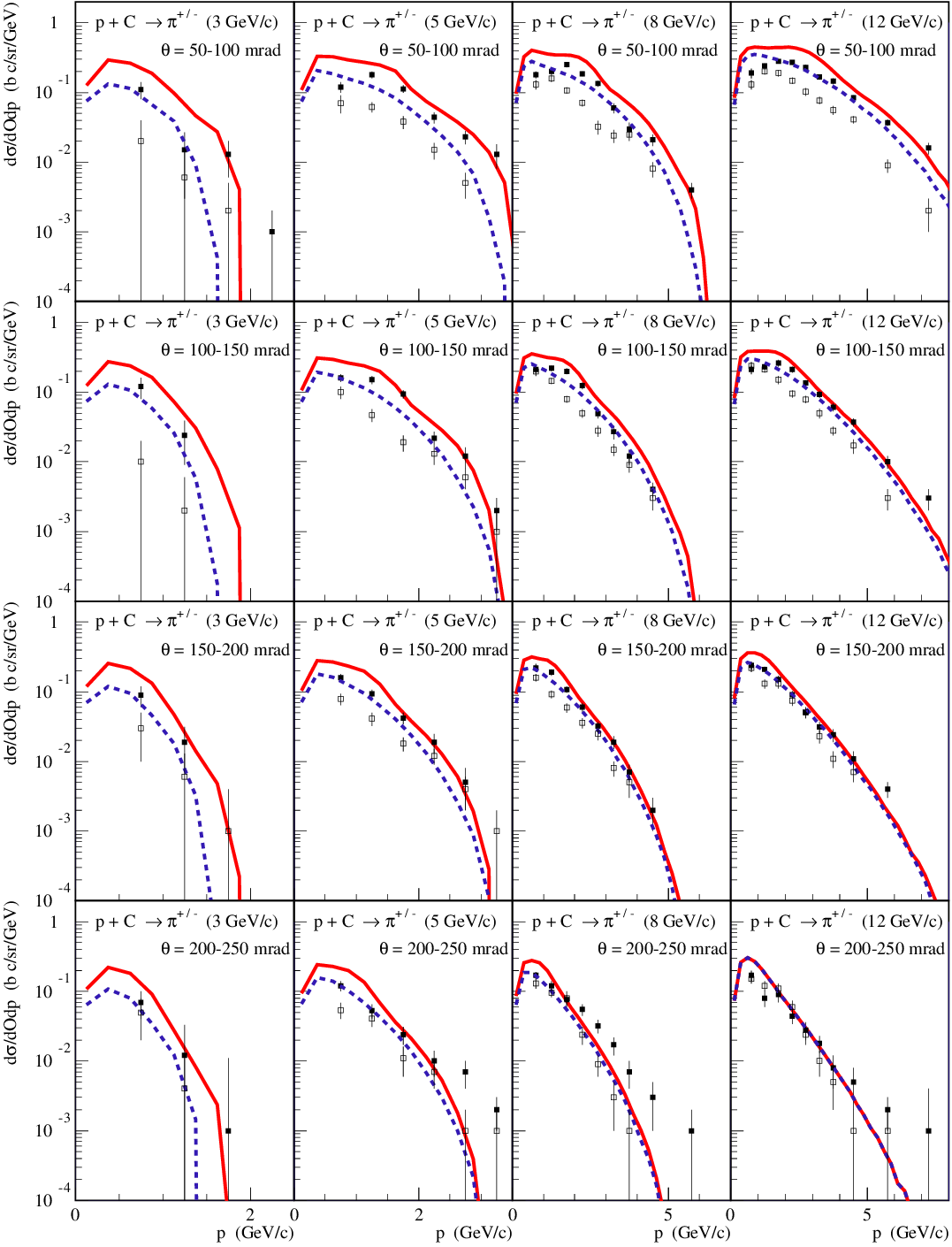}
\caption{Lab.\ momentum  dependence of inclusive cross sections for charged pion production, at  different polar angles,  in $p$C collisions at 3, 5, 8, and 12 GeV/c, as indicated in the plots, 
compared to HARP data \cite{apo09} (points):  $\pi^+$ -- red solid lines and filled  squares, $\pi^-$ -- blue dashed lines and open squares.}
\label{fig:pcharp}       % Give a unique label
\end{figure*}

\begin{figure*}[t]
\centering
\includegraphics[height=19cm,width=\textwidth]{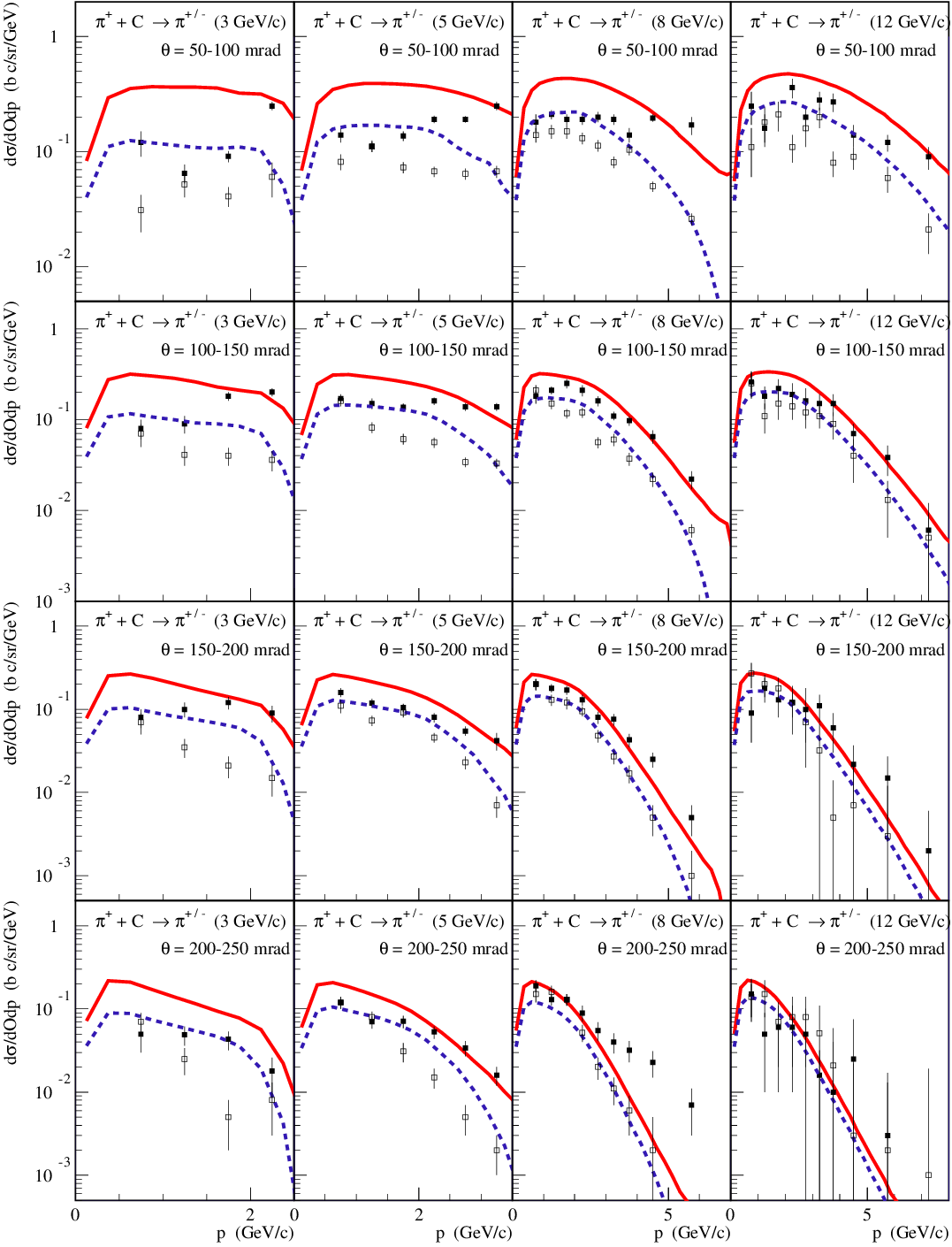}
\caption{Lab.\ momentum  dependence of inclusive cross sections for charged pion production, at  different polar angles,  in  $\pi^+$C collisions at 3, 5, 8, and 12 GeV/c, as indicated in the plots, 
compared to HARP data \cite{apo09a} (points):  $\pi^+$ -- red solid lines and filled  squares, $\pi^-$ -- blue dashed lines and open squares.}
\label{fig:pipcharp}       % Give a unique label
\end{figure*}

Going to even lower energies, we compare in Figs.\ \ref{fig:pcharp} and   \ref{fig:pipcharp} the
calculated  lab.\ momentum  dependence of inclusive cross sections for charged pion production in $p$C and $\pi^+$C collisions at 3, 5, 8, and 12 GeV/c, at  different polar angles,
 to the corresponding  data of the hadron production (HARP) experiment.
Here  the model results are again above  the  measured spectra at small angles,
the excess being higher at lower energies and in the target fragmentation region. Nevertheless, the overall description of forward pion production,
relevant for EAS modeling, remains reasonable down to very low energies.

Regarding the treatment of very high energy interactions dominated by semihard 
Pomeron exchanges, one has to take into consideration the changes in the underlying physics. Since the rapid energy rise of the semihard contribution is driven by copious production of hadron (mini)jets, the parameters governing the fragmentation of 
strings of color field, corresponding to   semihard Pomerons, should differ from the
ones for soft Pomerons, compiled in Table \ref{Flo:param-sstring}. First of all, as
discussed in Section \ref{physics.sec}, one expects in such a case a higher density
of produced hadrons per unit of rapidity, which can be achieved by using a larger
value for the parameter $\Lambda$ [cf.\ Eq.\ (\ref{fragm.eq})]. Additionally, the parameters  $\gamma_{qq}$,   $\gamma_{u}$, and $\gamma_{s}$, which govern the width
of transverse momentum distributions of various secondary hadrons, should have
larger values. Also the weights for diquark-antidiquark and $\bar ss$ pairs creation
from the vacuum may be different from the ones for soft Pomeron strings, reflecting
the differences in the kinematics regarding the hadronization of (mini)jets. 
In Table \ref{Flo:param-hstring}, we compile  
\begin{table*}[t]
\begin{centering}
\begin{tabular}{|lllll|}
\hline 
 $\Lambda \! ^{\rm (sh)}$ & $a_{ud}^{\rm (sh)}$ &    $\gamma_{qq}^{\rm (sh)}$ &    $\gamma_{u}^{\rm (sh)}=\gamma_{d}^{\rm (sh)}$  & $\gamma_{s}^{\rm (sh)}$
  \tabularnewline
 & &  {\scriptsize GeV$^2$}  &  {\scriptsize GeV$^2$}  &  {\scriptsize GeV$^2$} 
  \tabularnewline
\hline 
3.6 &  0.025  &  0.45 & 0.27 & 0.4
\tabularnewline
\hline
\end{tabular}\caption{String fragmentation parameters for semihard Pomerons, chosen
differently from the ``soft'' fragmentation  parameters.}
\label{Flo:param-hstring}
\par\end{centering}
 \end{table*}
 the parameters for fragmentation of   semihard Pomeron strings, which are chosen
 differently from the ``soft'' ones specified in  Table \ref{Flo:param-sstring}. 

\begin{figure}[htb]
\centering
\includegraphics[height=6.cm,width=\textwidth]{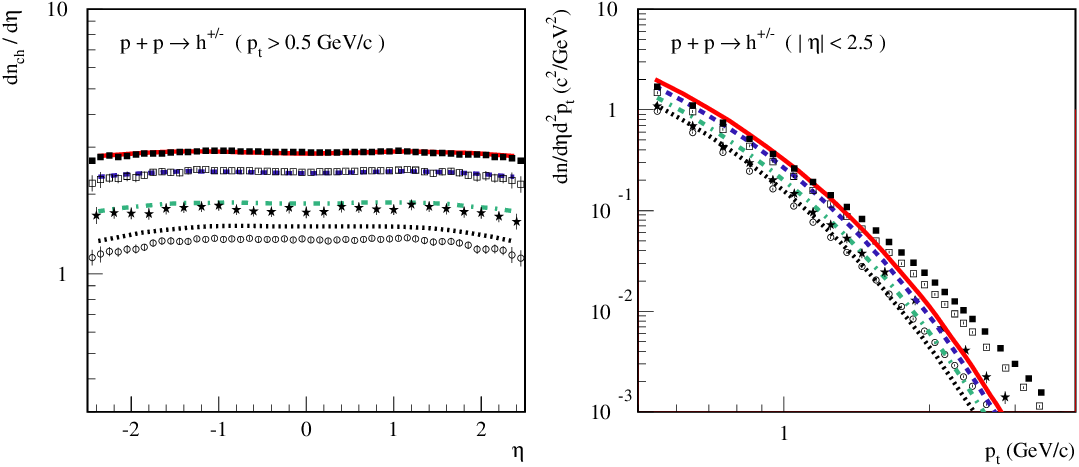}
\caption{Pseudorapidity (left) and transverse momentum (right) distributions of charged hadrons  in c.m.\ frame, 
produced in $pp$ collisions at different $\sqrt{s}$, 
compared to ATLAS data \cite{aad11,aad16} (points):
 $\sqrt{s}=13$ TeV -- red solid lines and filled squares;
 $\sqrt{s}=7$ TeV -- blue dashed lines and open squares;
$\sqrt{s}=2.36$ TeV -- green dash-dotted lines and filled stars;
$\sqrt{s}=0.9$ TeV --  black dotted lines and open circles.
 }
\label{fig:ppeta}       
\end{figure}

In Fig.\ \ref{fig:ppeta} (left), the calculated pseudorapidity $\eta$ distributions of charged
hadrons, $dN^{\rm ch}_{pp}/d\eta$, with transverse momenta $p_t>0.5$ GeV,
produced in $pp$ interactions at $\sqrt{s}=0.9$, 2.36, 7, and 13 TeV, are compared
to the corresponding data of the ATLAS experiment, for ATLAS event selection:
at least one charged hadron with  $p_t>0.5$ GeV in the $|\eta|<2.5$ range.
In turn, in Fig.\ \ref{fig:ppeta} (right), the corresponding  transverse momentum distributions,
$dN^{\rm ch}_{pp}/dp_t$, are shown. Overall, we have a reasonable agreement
with the measurements for  $dN^{\rm ch}_{pp}/d\eta$, over the considered energy range.
On the other hand, without an explicit treatment of (mini)jet production, the
model clearly fails to describe   high $p_t$ tails
 of the $dN^{\rm ch}_{pp}/dp_t$ distributions.\footnote{Such high  $p_t$ tails are irrelevant to EAS modeling since   very high energy parts of hadronic cascades in the
  atmosphere are strongly collimated, with hadron production angles in the lab.\
  frame, $\theta \simeq p_t/E$, being vanishingly small.}
  
\begin{figure*}[t]
\centering
\includegraphics[height=4.5cm,width=\textwidth]{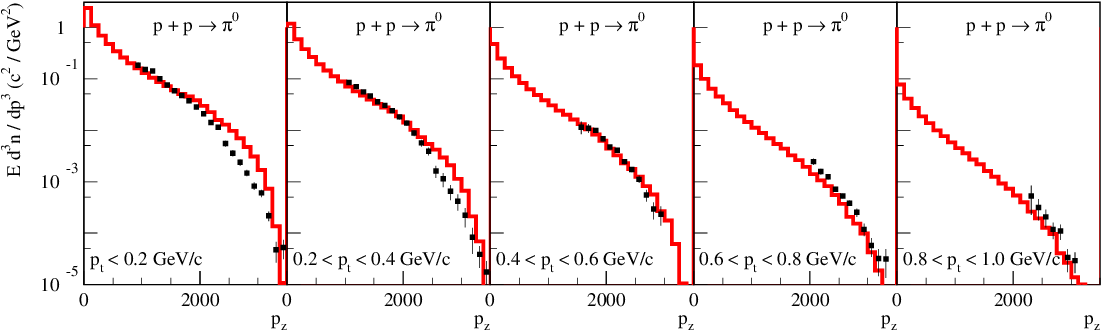}
\caption{Invariant momentum spectra of neutral pions in  c.m.~frame,
$E\, d^3n_{pp}^{\pi^{0}}/dp^3$, for  $pp$ collisions at $\sqrt{s}=7$ TeV, 
for different $p_t$ selections, as indicated
in the plots, compared to  LHCf data  \cite{adr16} (points).}
\label{Flo:lhcf-pi0}      
\end{figure*}

\begin{figure*}[t]
\centering
\includegraphics[height=11.cm,width=\textwidth]{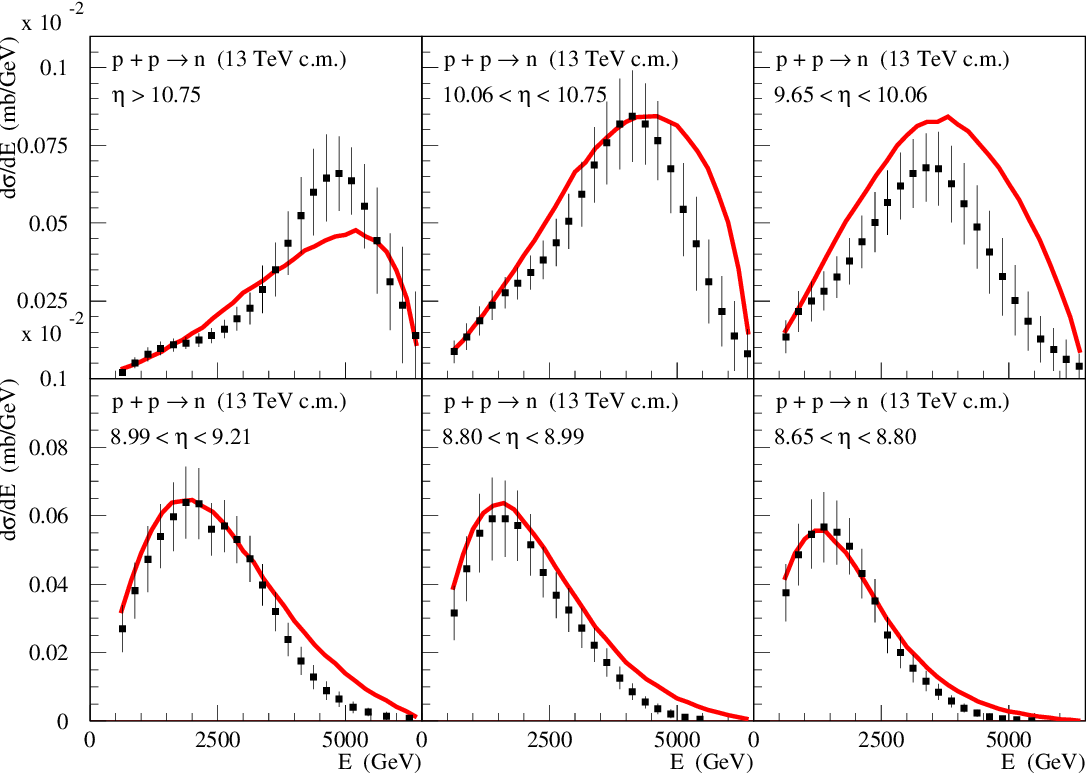}
\caption{Energy dependence of neutron production spectra in c.m.\ frame,
for  $pp$ collisions at $\sqrt{s}=13$ TeV,
for different pseudorapidity intervals, as indicated
in the plots, in comparison to LHCf data \cite{adr18} (points).}
\label{Flo:lhcf-n}       
\end{figure*}

  Regarding   model applications to EAS modeling, of major importance
  is an accurate description of forward particle production in hadronic collisions.
  In that regard, of particular value are measurements of forward production spectra
  of neutral pions and neutrons by the LHCf experiment \cite{adr16,adr18} ,
   our results being compared
  to the data in Figs.\ \ref{Flo:lhcf-pi0} and \ref{Flo:lhcf-n}.
  As one can see  in Fig.\ \ref{Flo:lhcf-pi0}, we have a good agreement with the
 measured spectra of neutral pions, for pion energies $E\lesssim 2$ TeV,
  corresponding to Feynman   $x_{\rm F}\lesssim 0.6$,
  while predicting harder pion spectra at larger  $x_{\rm F}$.\footnote{Such high
    $x_{\rm F}$ tails of secondary pion distributions
    are of minor importance for  EAS characteristics \cite{ost24d}.}
     Regarding our results for forward neutron spectra, shown in  
    Fig.\, \ref{Flo:lhcf-n}, we   have a reasonable
    overall agreement with the LHCf data. However, somewhat higher neutron 
    yields obtained by us   for neutron energies $E\gtrsim 4$ TeV  may indicate
    that the model underestimates the energy rise of the inelasticity of $pp$
     interactions, which, in turn, may lead to an underestimation of the speed
     of energy dissipation   in hadronic cascades in the atmosphere.

\section{Predictions for EAS characteristics \label{eas.sec}}
\begin{figure*}[t]
\centering
\includegraphics[height=6.cm,width=\textwidth]{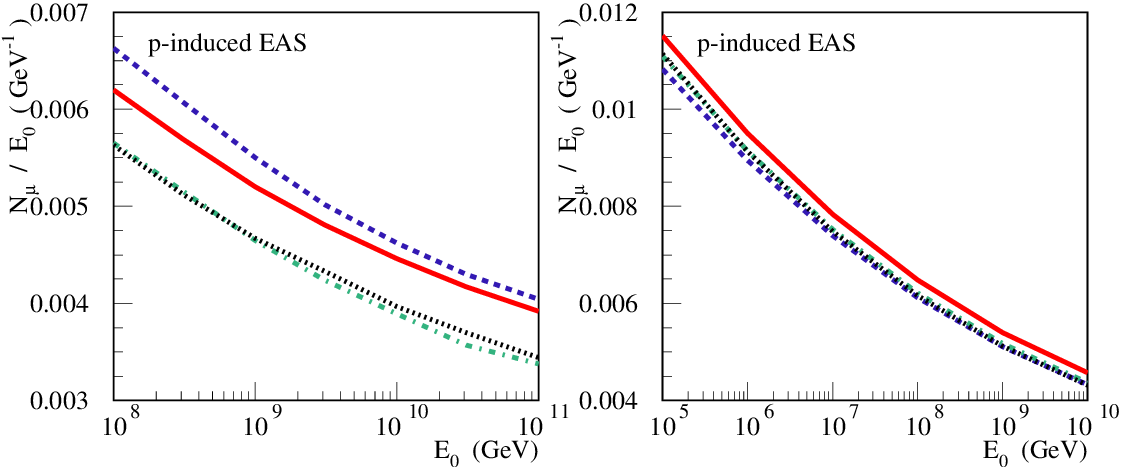}
\caption{Dependence on primary energy of   muon number  $N_{\mu}$ ($E_{\mu}>1$ GeV)
 at sea level, for proton-initiated vertical EAS,
 as calculated using  the QGSb model -- red solid lines. The corresponding results
 obtained using the EPOS-LHC-R, QGSJET-III, and  QGSJET-II-04 models for treating
 hadronic interactions above 80 GeV lab., while employing the  UrQMD  model at
 lower energies, are shown in the left panel by  blue dashed, green  dash-dotted,
  and black dotted lines, respectively. The results obtained using  QGSb above 80 GeV lab.\ and employing the FLUKA,  UrQMD, and GHEISHA models at lower energies are shown in the right panel by  blue dashed, green  dash-dotted,  and black dotted lines, respectively.}
\label{Flo:nmu}       
\end{figure*}

The developed model has been implemented in the CONEX EAS simulation 
program \cite{ber07} and applied for calculations of basic characteristics of CR-induced extensive air showers. In Fig.\ \ref{Flo:nmu}~(left), our results for the energy dependence of the number of muons $N_{\mu}$
at sea level, for muon energy $E_{\mu}>1$ GeV, in vertical proton-initiated EAS,
are compared to the corresponding predictions of the EPOS-LHC-R \cite{pie25},    QGSJET-III \cite{ost24b}, and QGSJET-II-04 \cite{ost11}
 CR interaction models.\footnote{In the corresponding calculations, 
 the QGSb model is used to treat hadronic 
interactions over the full relevant energy range. On the other hand,
the  EPOS-LHC-R, QGSJET-III, and QGSJET-II-04 models are applied to describe
hadron-air interactions above 80 GeV lab., while employing the
  UrQMD \cite{fes85}  model at lower energies.}
The values of $N_{\mu}$, obtained with QGSb, are relatively high,  approaching
at the highest CR energies the
corresponding results of the  EPOS-LHC-R MC generator. This is not surprising,
given a significant dependence of EAS muon content on forward production of
(anti)baryons in pion-air collisions \cite{pie08} and our calibration to the
respective results of the  NA61/SHINE experiment, rather than using alternative data sets (see the corresponding discussion in Section~\ref{results.sec}).
In  Fig.\ \ref{Flo:nmu}~(right), we compare  the energy dependence of $N_{\mu}$,
predicted by QGSb, to results of calculations using QGSb in the ``standard''
high energy range only, above 80 GeV lab., while treating hadron-air interactions 
at lower energies by means of the  FLUKA \cite{bat07},  UrQMD \cite{ble99}, and GHEISHA  models. As one can see in the Figure, using the QGSb MC generator for treating
the low energy part of extensive air showers gives rise to $\simeq 5$\% enhancement 
of  $N_{\mu}$, compared to calculations based on the traditional low energy
interaction models. This is again a consequence of the copious (anti)nucleon 
production in QGSb, which is related to the model calibration to the 
relevant data of NA61/SHINE. 

\begin{figure*}[t]
\centering
\includegraphics[height=6.cm,width=\textwidth]{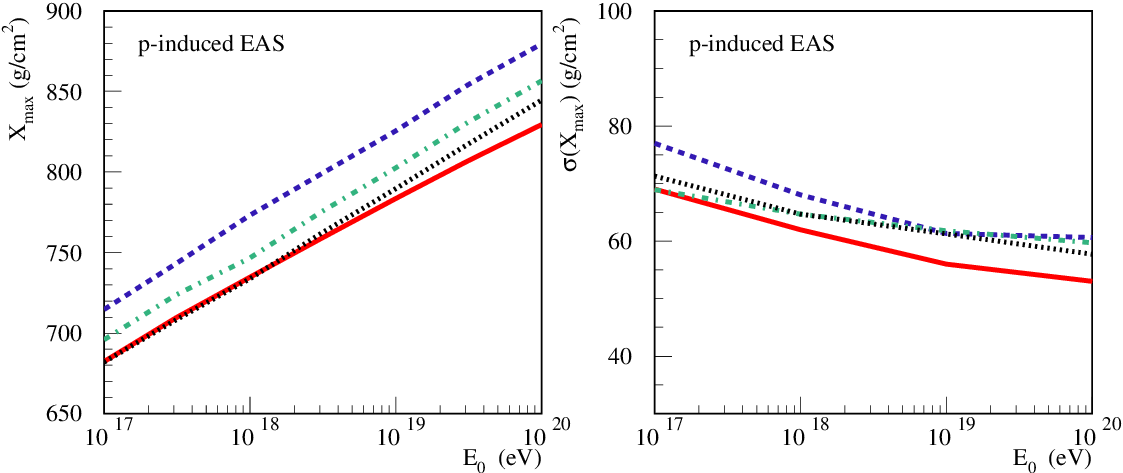}
\caption{Dependence on primary energy of    maximum depth $X_{\max}$ (left)
and of its fluctuations  $\sigma(X_{\max})$ (right),
 for proton-initiated vertical EAS,
 as calculated using  the QGSb model -- red solid lines.
 Blue dashed, green  dash-dotted, and black dotted lines correspond to
calculations  using the EPOS-LHC-R, QGSJET-III, and  QGSJET-II-04 models,
respectively, for treating
 hadronic interactions above 80 GeV lab., while employing the  UrQMD  model at
 lower energies.}
\label{fig:xmax}       
\end{figure*}

Further, in Fig.\ \ref{fig:xmax}, our predictions for EAS   maximum depth $X_{\max}$ 
and for the corresponding fluctuations are compared to the results of the
 EPOS-LHC-R,   QGSJET-III, and QGSJET-II-04 CR interaction models. 
  Interestingly, at the highest energies,
 the obtained elongation
rate of extensive air showers, $d\lg X_{\max}(E_0)/dE_0$, is noticeably
smaller, compared to earlier predictions. This is partly related to our
treatment of the pion exchange process in proton-proton and proton-nucleus collisions, 
which takes into consideration the contribution of elastic scattering of virtual pions. As demonstrated in 
Fig.\ \ref{fig:pdiffrt}, the process contributes significantly to the 
 peaks in proton spectra  at $|x_{\rm F}|\simeq 1$
  and thereby impacts the
calibration of low mass diffraction treatment of the model to the
relevant experimental data. With increasing energy, the pion exchange is
pushed towards large impact parameters, which leads to a significant
suppression of the  contribution of elastic scattering of virtual pions, hence,
to a noticeable increase of the inelasticity of  proton-proton and proton-nucleus collisions. Another reason for a somewhat higher inelasticity
of our model is our treatment of Reggeon exchange contributions, which impacts the model calibration at fixed target energies. In proton-proton and proton-nucleus collisions, the ``undeveloped cylinder'' process 
exemplified in Fig.\ \ref{fig:planar} (right) contributes noticeably to forward
production of nucleons. However, with increasing energy, the process quickly
dies out, which leads effectively to an increasing energy loss of leading nucleons. Yet   predictions for  $X_{\max}$ depend strongly on model assumptions
regarding momentum distributions of constituent partons to which string of color field are attached \cite{ost24c,ost03,par11,ost16}, which is controlled by the
$\alpha _{\rm sea}$ parameter of the model. Choosing a larger value for $\alpha _{\rm sea}$  would shift our predictions for EAS maximum depth   deeper in the atmosphere. The corresponding model uncertainty will be studied elsewhere.

Regarding fluctuations of  EAS maximum depth, our results for 
 $\sigma(X_{\max})$, plotted in  Fig.\ \ref{fig:xmax} (right),
  agree  with the ones of the other models. This is not surprising, given
  very precise measurements of total and elastic proton-proton interaction cross sections  at the Large Hadron Collider. The spread of the model predictions for  $\sigma(X_{\max})$
   can be explained by uncertainties regarding the rate of inelastic diffraction in $pp$ collisions \cite{ost14}.

\begin{figure}[htb]
\centering
\includegraphics[height=6.cm,width=0.49\textwidth]{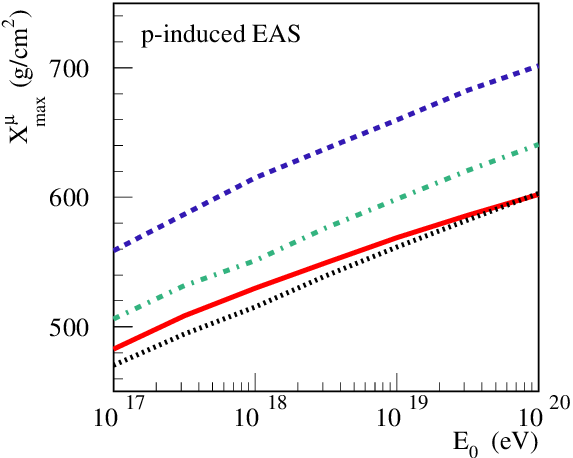}
\caption{Dependence on primary energy of  maximal muon
production depth $X^{\mu}_{\max}$ ($E_{\mu}>1$ GeV),
 for proton-initiated vertical EAS,
 as calculated using  the QGSb model -- red solid line.
 Blue dashed, green  dash-dotted, and black dotted lines correspond to
calculations  using the EPOS-LHC-R, QGSJET-III, and  QGSJET-II-04 models,
respectively, for treating
 hadronic interactions above 80 GeV lab., while employing the  UrQMD  model at
 lower energies.}
\label{fig:xmumax}       
\end{figure}

Finally, our predictions for maximal muon
production depth $X^{\mu}_{\max}$  
are compared to the corresponding results of  the other
   CR interaction models  in  Fig.\ \ref{fig:xmumax}.
 As discussed in \cite{ost16a}, measurements of this quantity by the Pierre Auger
 Observatory \cite{aab14} allow one to substantially constrain both the
 treatment of pion-air interactions at the highest energies and the 
 corresponding predictions for   $X_{\max}$.
 The obtained  values of $X^{\mu}_{\max}$ are  between the results
 of the other models, approaching the ones of the  QGSJET-II-04 MC generator
 at the highest energies. The origin of the latter trend is the same as for
   $X_{\max}$ above: being mostly related to the contribution of elastic
   scattering of virtual pions in the pion exchange process, dying out with energy.

\section{Outlook \label{concl.sec}}
We  developed  a new  MC generator designed  to treat hadronic interactions
in a broad energy range: from $\sim 10$ GeV lab.\ 
up to the highest energies of cosmic rays observed.
We relied on a relatively simple and transparent interaction picture,
 while using the theoretical formalism of the Reggeon Field Theory \cite{gri68,gri69}.
 Basically, we employed the  concepts of
the original Quark-Gluon String and Dual Parton models \cite{kai82,cap91},
describing high energy hadron-proton, hadron-nucleus, and nucleus-nucleus collisions
by multi-Pomeron exchanges.

To mimic the transition from soft to semihard physics, regarding the
 ``parton content'' of the Pomeron, we   considered two contributions
to the Pomeron amplitude: the one corresponding to  soft 
parton cascade, characterized by a rather slow energy rise, and the  semihard 
one quickly rising with energy.

Regarding the description of inelastic diffraction, we relied on the traditional
 Good-Walker  approach to treat low mass diffractive excitations of interacting
 hadrons (nucleons) and the corresponding inelastic screening effects.
 In turn,  for high mass diffraction,  we considered the diffractive cut of 
 the simplest triple-Pomeron  graph 
 corresponding to an interaction between three  soft  Pomerons.

In view of its importance for EAS muon number predictions,
 we took into consideration the  $t$-channel pion exchange  process,
   improving the respective treatment of the  QGSJET-III model.  
 Namely, we considered explicitly a rescattering of  virtual pions
 emitted by interacting hadrons (nucleons),
including the corresponding inelastic and elastic scattering contributions and 
accounting for   impact parameter dependent absorptive corrections for the process.

In order to extend the model for a description of hadronic interactions at 
relatively low energies, below 100 GeV lab.,
 we took into account  contributions of secondary Reggeon exchanges,
 following the approach proposed in  \cite{don92}:
 considering an exchange of a single effective Reggeon per reaction type.

The parameters of the model have been tuned based on a variety of accelerator data
regarding  the energy dependence of total, elastic, and particle production
cross sections for hadron-proton and hadron-nucleus collisions, and concerning
various characteristics of secondary particle production. Overall, the relatively
high parameter freedom of the model allowed us to achieve a satisfactory description
of the relevant experimental data over the wide energy range considered.

The developed model has been  applied for calculations of basic characteristics 
of CR-induced extensive air showers:  number of muons at ground level,
 EAS   maximum depth   and its fluctuations,  and  maximal muon production depth.
The obtained results appeared to be consistent with the corresponding predictions 
 of the standard  CR interaction models. Yet, at the highest CR energies,
 the EAS elongation rate predicted by the new model is noticeably
smaller, compared to earlier predictions. This stresses the need to consider
alternative tunes of the model parameters, allowed by accelerator data,
and to study the corresponding uncertainties for EAS predictions.
Such an investigation will be reported elsewhere.

One important remark is in order here. The presented model provides an effective
decription of high energy cosmic ray interactions, addressing the basic physics
which is of relevance for calculations of extensive air shower development.
However, the latter can potentially be influenced by some other, more specific,
interaction mechanisms, e.g., ones discussed in \cite{pie25,dre05}, which may be
implemented at later stages.

Generally, the developed MC generator can be used both for  
studying model uncertainties for  EAS predictions and for investigating
a possibility to describe particular sets of CR data by suitably retuning the
model parameters. It is noteworthy that a high computational efficiency
of the developed model may allow one to apply automated tuning procedures,
e.g., using a combination of accelerator and CR data, as envisaged in \cite{alb25}.

Apart from EAS applications,  the developed MC generator can be applied to
direct CR studies, e.g., regarding calculations of electron, positron, photon,
and (nonprompt) neutrino production in interactions of high energy cosmic rays with the interstellar medium. In that respect, it may complement existing parametrizations of secondary particle spectra for proton-proton and proton-nucleus collisions \cite{kel06,kam06,kos18,kac19,oru22,oru23,oru26} since the efficiency and transparency of the model can offer here the same advantages as for EAS  applications. Namely, potential users of the model may attempt to perform their own tuning of its parameters, either aiming at improving the agreement with particular experimental data sets or investigating the impact of uncertainties regarding the treatment of hadronic interactions on interpretations of various astrophysical data sets.

\subsection*{Acknowledgments}
S.O.\ acknowledges   support from  Deutsche Forschungsgemeinschaft 
(project  550225003).  G.S.\ acknowledges
support by the Bundesministerium f\"ur Bildung
und Forschung, under grant 05A23GU3, and
by the Deutsche Forschungsgemeinschaft  under    
  Germany's Excellence Strategy -- EXC 2121 ``Quantum Universe'' -- 390833306.
 This research was supported in part by the
Munich Institute for Astro-, Particle and BioPhysics (MIAPbP) which is funded by the Deutsche
Forschungsgemeinschaft (DFG, German Research Foundation) under Germany's Excellence
 Strategy -- EXC-2094 -- 390783311. The authors acknowledge useful discussions with   T.\ Sj\"ostrand and other participants
of the 2025 MIAPbP program ``Event Generators at Colliders and Beyond Colliders''.


\begin{thebibliography}{99}

  \bibitem{nag00}
  M.\ Nagano and A.\ A.\ Watson, 
  {\em Observations and implications of the ultrahigh-energy cosmic rays}, 
   Rev.\ Mod.\ Phys.\ {\bf 72}, 689 (2000).
   
  \bibitem{blu09}  
  J.\ Bl\"umer, R.\ Engel, and  J.\ R.\ H\"orandel,
{\em Cosmic rays from the knee to the highest energies}, 
        Prog.\ Part.\ Nucl.\ Phys.\  {\bf  63}, 293 (2009).

\bibitem{eng11}
 R.\ Engel, D.\ Heck, and T.\ Pierog, 
{\em  Extensive air showers and hadronic interactions at high energy}, 
 Ann.\ Rev.\ Nucl.\ Part.\ Sci.\  {\bf 61}, 467 (2011).

\bibitem{pie15}
   T.\ Pierog, Iu.\ Karpenko, J.\ M.\ Katzy, E.\ Yatsenko, and K.\ Werner,
{\em  EPOS LHC: Test of collective hadronization with data 
measured at the CERN Large Hadron Collider},
   Phys.\  Rev.\ C {\bf 92},  034906     (2015).  

\bibitem{rie20}
  F.\ Riehn, R.\ Engel, A.\ Fedynitch, T.\ K.\ Gaisser, and T.\ Stanev,
{\em  Hadronic interaction model Sibyll 2.3d and extensive air showers},
   Phys.\  Rev.\ D {\bf 102},  063002     (2020).  
   
   \bibitem{ost24a} S.\ Ostapchenko, 
 {\em QGSJET-III model of high energy hadronic interactions: The formalism},
  Phys.\  Rev.\ D {\bf  109},   034002 (2024).  
 
  \bibitem{abr13}
 P.\ Abreu  et al.\ (Pierre Auger Collaboration), 
 {\em   Interpretation of the depths of maximum of
extensive air showers measured by the Pierre Auger Observatory}, 
   JCAP   {\bf 02}, 026 (2013).

  \bibitem{aab14}
A.\ Aab et al.\ (Pierre Auger Collaboration), 
{\em  Muons in air showers at the Pierre Auger Observatory:
 Measurement of atmospheric production depth}, 
 Phys.\  Rev.\ D {\bf  90},   012012  (2014).  

 \bibitem{aab15}
A.\ Aab et al.\ (Pierre Auger Collaboration), 
{\em Muons in air showers at the Pierre Auger
Observatory: Mean number in highly inclined events}, 
 Phys.\  Rev.\ D {\bf  91},   032003   (2015).  

   \bibitem{aab16}
  A.\ Aab et al.\ (Pierre Auger Collaboration), 
   {\em   Testing hadronic interactions at ultrahigh
  energies with air showers measured by the Pierre Auger Observatory}, 
      Phys.\ Rev.\ Lett.~{\bf 117}, 192001   (2016).	

 \bibitem{abd24}
 A.\  Abdul Halim  et al.\ (Pierre Auger Collaboration), 
{\em Testing hadronic-model predictions of depth of maximum of air-shower profiles and 
ground-particle signals using hybrid data of the Pierre Auger Observatory},
 Phys.\  Rev.\ D {\bf 109},   102001    (2024).  

 \bibitem{ebr23}
 J.\ Ebr, J.\ Bla\^zek, J.\ V\'icha, T.\ Pierog,  E.\ Santos, P.\ Tr\'avn\'i\^cek,   N.\ Denner,  and R.\ Ulrich,
{\em  Impact of modified characteristics of hadronic interactions on cosmic-ray observables for proton and nuclear primaries},
PoS \textbf{ICRC2023},  245 (2023).

 \bibitem{gri68}
 V.~N.~Gribov,  
{\em   A reggeon diagram technique}, 
 Sov.~Phys.~JETP~{\bf  26}, 414 (1968). 
 
\bibitem{gri69}
 V.\ N.\ Gribov, 
 {\em Glauber corrections and the interaction between high-energy 
hadrons and nuclei},
  Sov.\ Phys.\ JETP  {\bf 29}, 483, (1969). 
  
  \bibitem{kai82} 
A.~B.~Kaidalov and K.~A.~Ter-Martirosyan, 
 {\em Pomeron as quark-gluon strings and multiple hadron 
production at SPS collider energies},
 Phys.~Lett.~B   {\bf 117},  247 (1982).

\bibitem{cap91} 
A.~Capella,  U.\ Sukhatme, C.-I.\ Tan, and J.~Tran Thanh Van,
{\em Dual parton model},
 Phys.\ Rep.\   {\bf 236},  225 (1994).

\bibitem{don92} 
   A.\ Donnachie  and  P.\ V.\  Landshoff,
 {\em  Total cross-sections},
       Phys.\ Lett.\ B {\bf 296}, 227 (1992).

\bibitem{kal93}
N.~N.~Kalmykov and  S.~S.~Ostapchenko,
 {\em  The nucleus-nucleus interaction, nuclear fragmentation, and fluctuations
 of extensive air showers},  
 Phys.~Atom.~Nucl.~{\bf 56}, 346 (1993).
	
\bibitem{ost11} S. Ostapchenko, 
 {\em Monte Carlo treatment of hadronic 
interactions in enhanced pomeron scheme: QGSJET-II model}, 
 Phys.\  Rev.\ D {\bf 83},  014018 (2011).

   \bibitem{ost24b} S.\ Ostapchenko, 
 {\em QGSJET-III model of high energy hadronic interactions: II. Particle
production and extensive air shower characteristics},
 Phys.\  Rev.\ D {\bf  109},   094019  (2024).  

\bibitem{glr} L.~Gribov, E.~Levin, and M.~Ryskin,
{\em Semihard processes in QCD},
Phys.\ Rep.\  {\bf 100}, 1 (1983).

\bibitem{don98} 
A.\ Donnachie and P.\ V.\ Landshoff, 
{\em Small x: Two pomerons!}, 
 Phys.\ Lett.\ B  {\bf 437}, 408 (1998).

\bibitem{ost02}
 S.\  Ostapchenko,  H.\ J.\ Drescher,  F.\ M.\  Liu, T.\ Pierog,  and K.\  Werner,  
{\em Consistent treatment of soft and hard processes in hadronic interactions},
  J.\ Phys.\ G  {\bf 28}, 2597 (2002).

\bibitem{ost06} 
S.~Ostapchenko,
  {\em On the re-summation of enhanced pomeron diagrams},
 Phys.\ Lett.\ B  {\bf 636},   40 (2006).

\bibitem{kai86}  
A.~B.~Kaidalov, L.~A.~Ponomarev, and K.~A.~Ter-Martirosyan, 
 {\em  Total cross-sections and diffractive scattering in a theory of interacting pomerons with $\alpha_P(0) > 1$},
  Sov.~J.~Nucl.~Phys.~{\bf 44}, 468 (1986).
 
\bibitem{col77} 
P.\ D.\ B.\ Collins,
 {\em An introduction to Regge theory and high energy physics},
 Cambridge U.P., 1977.

 \bibitem{fra08}
 L.~Frankfurt, M.~Strikman, D.~Treleani, and C.~Weiss,
    {\em Evidence for color fluctuations in the nucleon in high-energy scattering},
     Phys.\ Rev.\ Lett.~{\bf 101}, 202003 (2008).	
 
\bibitem{goo60}
M.\ L.\ Good and W.\ D.\ Walker,
  {\em Diffraction disssociation of beam particles},
 Phys.\ Rev.\ {\bf 120}, 1857 (1960).

\bibitem{ost10} S.\ Ostapchenko,
  {\em Total and diffractive cross sections in enhanced pomeron scheme},
 Phys.~Rev.~D    {\bf  81}, 114028 (2010). 

     \bibitem{agk74}
 V.\ A.\ Abramovsky,  V.\ N.\ Gribov, and  O.~V.~Kancheli, 
  {\em Character of inclusive spectra and fluctuations produced in
  inelastic processes by multi-pomeron exchange}, 
  Sov.~J.~Nucl.~Phys.\   {\bf   18}, 308 (1974).
 
\bibitem{kai87}
A.~B.~Kaidalov,  
 {\em Quark and diquark fragmentation functions in the model of quark gluon strings},
  Sov.~J.~Nucl.~Phys.~{\bf 45}, 902 (1987).
 
\bibitem{eng91}
  J.\ Engel,  T.\ K.\ Gaisser, T.\ Stanev, and P.\ Lipari,
{\em  Nucleus-nucleus collisions and interpretation of cosmic ray cascades},
   Phys.\  Rev.\ D {\bf 46},  5013  (1992).  

  \bibitem{zol88}
    V.\ R.\  Zoller,
    {\em  Diffractive scattering off nuclei in the multiple scattering theory with inelastic screening},
  Sov.~J.~Nucl.~Phys.~{\bf 48}, 361(1988).

	\bibitem{pdg}
 R.\ L.\ Workman  {\em et al.} (Particle Data Group), 
 {\em Review of Particle Physics},
   Prog.\ Theor.\ Exp.\ Phys.\  {\bf 2022}, 083C01 (2022).
    

   \bibitem{ost24c} S.\ Ostapchenko and G.\ Sigl, 
   {\em Model uncertainties for the predicted maximum depth of extensive air showers},
 Phys.\  Rev.\ D {\bf  110},   063041   (2024).  

  \bibitem{ost13}
S.~Ostapchenko,
   {\em QGSJET-II: physics, recent improvements, and results for
  air showers},
  EPJ Web Conf.\   {\bf 52},    02001 (2013).

  \bibitem{ost21}
  S.~Ostapchenko,
    {\em QGSJET-III model: novel features},
   Phys.\ At.\ Nucl.\   {\bf 44},  1017 (2021).

 \bibitem{kai06}
 A.~B.~Kaidalov, V.\ A.\ Khoze, A.\ D.\ Martin, and M.\ G.\ Ryskin,  
   {\em  Leading neutron spectra},
         Eur.\ Phys.\ J.\ C {\bf 47}, 385  (2006).

\bibitem{ant19}
 G.\ Antchev et al. (TOTEM Collaboration),
{\em First measurement of elastic, inelastic and total cross-section 
at  $\sqrt{s}=13$ TeV by TOTEM and overview of cross-section data at LHC energies},	
 Eur.\ Phys.\ J.\  C \textbf{79},  103 (2019).

\bibitem{aad23}
%G.\ Aad, B.\ Abbott,  D.\ C.\ Abbott  et al.\ (ATLAS Collaboration),
G.\ Aad  et al.\ (ATLAS Collaboration),
  {\em  Measurement of the total cross section and $\rho$-parameter from elastic scattering in pp collisions at $\sqrt{s}=13$ TeV with the ATLAS detector},
 Eur.~Phys.~J.~C {\bf 83},  441  (2023).

   \bibitem{ost19}
 S.~Ostapchenko and   M.~Bleicher, 
   {\em Taming the energy rise of the total proton-proton cross-section},
  Universe  {\bf 5},   106 (2019).

   \bibitem{bel66}
 G.\  Bellettini, G.\  Cocconi, A.\ N.\  Diddens, E.\  Lillethun, G.\  Matthiae, J.\ P.\  Scanlon,
 and  A.~M.~Wetherell,
 {\em  Proton-nuclei cross sections at 20 GeV},
  Nucl.\ Phys.\    {\bf  79},   609 (1966).
 
   \bibitem{eng70}
    J.\ Engler et al.,
{\em   Neutron-nucleus total cross-sections between 8 GeV/c and 21 GeV/c},
Phys.\ Lett.\ B  {\bf 32},  716 (1970).
 
   \bibitem{jon71}
   L.\  W.\   Jones, M.\  J.\   Longo, T.\  P.\   McCorriston, E.\  F.\   Parker, S.\  T.\   Powell, 
   and M.\  N.\   Kreisler,
  {\em   Neutron total cross-sections on protons and nuclei in the 
  10 to 30 GeV/c momentum range},
 Phys.\ Lett.\ B  {\bf 36},   509 (1971).

   \bibitem{den73}
    S.\ P.\  Denisov,  S.\  V.\   Donskov, Yu.\  P.\   Gorin, R.\  N.\   Krasnokutsky, A.\  I.\   Petrukhin, Yu.~D.~Prokoshkin, and D.\  A.\   Stoyanova,
  {\em  Absorption cross-sections for pions, kaons, protons and anti-protons 
  on complex nuclei in the 6  to 60 GeV/c momentum range},
 Nucl.\ Phys.\ B  {\bf 61},   62 (1973).
  
   \bibitem{bab74}
    A.\  Babaev et al.,
 {\em Total cross-section measurement of neutrons on protons and nuclei over
  the energy range 28--54 GeV},
 Phys.\ Lett.\ B  {\bf 51},  501 (1974).
  
   \bibitem{clo74}
    A.\ S.\  Clough et al.,
 {\em Pion-nucleus total cross-sections from 88  to 860 MeV},
 Nucl.\ Phys.\ B  {\bf 76},  15 (1974).

\bibitem{mur75}
    P.\ V.\ R.\ Murthy, C.\ A.\ Ayre, H.\ R.\ Gustafson, L.\ W.\ Jones, and  M.\ J.\ Longo,
{\em Neutron total cross sections on nuclei at Fermilab energies},	
 Nucl.\ Phys.\ B  {\bf 92}, 269 (1975).

\bibitem{car79}
    A.\ S.\  Carroll et al., 
{\em Total cross-Sections of $\pi^{\pm}$,  $K^{\pm}$, $p$ and $\bar p$ on 
protons and deuterons between 200 GeV/c and 370 GeV/c},
 Phys.\ Lett.\ B  {\bf 80},   423 (1979).

\bibitem{der00}
 U.\ Dersch et al.\ (SELEX Collaboration),
{\em Total cross-section measurements with $\pi^-$,  $\Sigma^-$ and protons on nuclei around 600 GeV/c},	
 Nucl.\ Phys.\ B  {\bf 579},  277 (2000).
 
\bibitem{adu18}
    A.\ Aduszkiewicz et al. (NA61/SHINE Collaboration),
{\em Measurements of total production cross sections for $\pi^+$+C, $\pi^+$+Al,
 $K^+$+C, and $K^+$+Al at 60 GeV/c and  $\pi^+$+C and $\pi^+$+Al at 31 GeV/c},
 Phys.~Rev.~D    {\bf 98},   052001 (2018). 

 \bibitem{adu19}
    A.\ Aduszkiewicz et al. (NA61/SHINE Collaboration),
{\em Measurements of production and inelastic cross sections for p+C, p+Be,
 and p+Al at 60 GeV/c and p+C and p+Be at 120 GeV/c},
 Phys.~Rev.~D    {\bf 100},  112001 (2019). 

\bibitem{ach21}
  A. Acharya et al.\ (NA61/SHINE Collaboration),
{\em   Measurement of the production cross section of 31 GeV/c protons 
on carbon via beam attenuation in a 90-cm-long target},
 Phys.~Rev.~D    {\bf 103}, 012006 (2021). 
 
  \bibitem{gla56}
 R.\ J.\ Glauber, 
 {\em  High-energy collision theory},
 In: Lectures in theoretical physics, ed.\ by W.~E.~Brittin 
and L.\ G.\ Dunham, Interscience Publishers (New York, 1959), vol.\ 1,
 pp.\ 315-414.
   
\bibitem{alt06} 
C.\ Alt  {\em et al.} (NA49 Collaboration),
{\em Inclusive production of charged pions in p+p collisions at 
158 GeV/c beam momentum},
  Eur.\ Phys.\ J.\  C  {\bf 45},  343 (2006).

\bibitem{ant10} 
 T.\ Anticic  {\em et al.} (NA49 Collaboration),
{\em  Inclusive production of protons, anti-protons and 
neutrons in p+p collisions at 158-GeV/c beam momentum},
  Eur.\ Phys.\ J.\  C  {\bf 65},   9 (2010).
  
\bibitem{ant10a} 
 T.\ Anticic  {\em et al.} (NA49 Collaboration),
{\em Inclusive production of charged kaons in p+p collisions at
 158 GeV/c beam momentum and a new evaluation of the energy
  dependence of kaon production up to collider energies},
  Eur.\ Phys.\ J.\  C  {\bf 68},  1 (2010).
 
\bibitem{alt07} 
C.\ Alt  {\em et al.} (NA49 Collaboration),
{\em Inclusive production of charged pions in p+C collisions at
 158 GeV/c beam momentum},
  Eur.\ Phys.\ J.\  C  {\bf 49},  897 (2007).
 
\bibitem{baa13} 
  B.\ Baatar  {\em et al.} (NA49 Collaboration),
{\em  Inclusive production of protons, anti-protons, neutrons, deuterons and tritons in p+C collisions at 158 GeV/c beam momentum},
  Eur.\ Phys.\ J.\  C  {\bf 73},  2364 (2013).

\bibitem{adh23} 
H.\ Adhikary  {\em et al.} (NA61/SHINE Collaboration),
{\em Measurement of hadron production in $\pi^-$-C interactions at 158 and 350 GeV/c
 with NA61/SHINE at the CERN SPS},
  Phys.\ Rev.\  D   {\bf 107},  062004 (2023).
  
\bibitem{adh25} 
H.\ Adhikary  {\em et al.} (NA61/SHINE Collaboration),
{\em  Evidence of isospin-symmetry violation in high-energy collisions 
of atomic nuclei},
        Nature Commun.\   {\bf 16},  2849 (2025).

\bibitem{ost24d}
S.\ Ostapchenko and G.\ Sigl,
{\em  On the model uncertainties for the predicted muon
content of extensive air showers},
 Astropart.\ Phys.\ {\bf 163},  103004 (2024).

\bibitem{adu17} 
A.\ Aduszkiewicz  {\em et al.} (NA61/SHINE Collaboration),
{\em Measurement of meson resonance production in $\pi^-$-C interactions at SPS energies},
   Eur.\ Phys.\ J.\  C  {\bf 77}, 626 (2017).


\bibitem{whi75} 
  J.\ Whitmore, S.\ J.\  Barish, D.\  C.\   Colley, and  P.\ F.\  Schultz,
{\em Invariant cross-section for the inclusive reaction $p+p\rightarrow p+X$ at 205 GeV/c},
  Phys.~Rev.~D    {\bf 11}, 3124 (1975). 

\bibitem{aji89} 
    I.\ V.\ Ajinenko  {\em et al.} (EHS/NA22 Collaboration),
{\em Strange and nonstrange baryon production in $\pi^+p$  and  $K^+p$ interactions at 250 GeV/c},
        Z.\ Phys.\ C  {\bf 44}, 573 (1989).

 \bibitem{olj20}  
 F.~Oljemark,
{\em Single Diffraction in proton-proton scattering with TOTEM 
at the Large Hadron Collider}, PhD thesis,
University of Helsinki (2020).
	
\bibitem{adu17a} 
A.\ Aduszkiewicz  {\em et al.} (NA61/SHINE Collaboration),
{\em Measurements of  $\pi^{\pm}$,   $K^{\pm}$, $p$ and $\bar p$
spectra in proton-proton interactions at 20, 31, 40, 80 and 158  GeV/c with the NA61/SHINE spectrometer at the CERN SPS},
   Eur.\ Phys.\ J.\  C  {\bf 77}, 671 (2017).

\bibitem{adu19a} 
A.\ Aduszkiewicz  {\em et al.} (NA61/SHINE Collaboration),
{\em Measurements of hadron production in $\pi^+ + {\rm C}$ and
 $\pi^+ + {\rm Be}$  interactions at 60 GeV/c},
  Phys.~Rev.~D    {\bf 100},  112004 (2019). 
  
\bibitem{apo09} 
M. Apollonio  {\em et al.} (HARP Collaboration),
{\em Forward production of charged pions with incident protons 
on nuclear targets at the CERN PS},
  Phys.~Rev.~C    {\bf 80},  035208 (2009). 
 
\bibitem{apo09a} 
M. Apollonio  {\em et al.} (HARP Collaboration),
{\em Forward production of charged pions with incident $\pi^{\pm}$ 
on nuclear targets measured at the CERN PS},
 Nucl.\ Phys.\ A  {\bf 821},  118 (2009).
 
\bibitem{aad11}
G.\ Aad  et al.\ (ATLAS   Collaboration),
{\em  Charged-particle multiplicities in pp interactions measured with the ATLAS detector at the LHC},
	New J.\ Phys.\  \textbf{13}, 053033 (2011).
	
\bibitem{aad16}
G.\ Aad   {\em et al.} (ATLAS  Collaboration),
{\em Charged-particle distributions in $\sqrt{s}=13 $ TeV
$pp$ interactions measured with the ATLAS detector at the LHC},
 Phys.\ Lett.\ B \textbf{758},  67 (2016).

\bibitem{adr16}
O.~Adriani    {\em et al.}  (LHCf Collaboration),
{\em Measurements of longitudinal and transverse momentum distributions
for neutral pions in the forward-rapidity region with the LHCf detector},
 Phys.\ Rev.\  D  {\bf 94}, 032007 (2016).

   \bibitem{adr18} 
  O.\ Adriani   {\em et al.} (LHCf Collaboration),
  {\em Measurement of inclusive forward neutron production cross 
   section in proton-proton collisions at $\sqrt{s}=13$ TeV with
    the LHCf Arm2 detector},
   J.~High Energy Phys.\       {\bf 11}, 073 (2018). 

\bibitem{ber07}
T.\ Bergmann, R.\ Engel, D.\ Heck, N.\ N.\ Kalmykov, S.~Ostapchenko,
 T.\ Pierog, T.\ Thouw, and K.\ Werner,
 {\em  One-dimensional hybrid approach to extensive air shower simulation},
  Astropart.\ Phys.\   {\bf 26}, 420 (2007).
 
  \bibitem{pie25}
   T.\ Pierog and K.\ Werner,
 {\em EPOS LHC-R : a global approach to solve the muon puzzle},
PoS \textbf{ICRC2025}, 358 (2025).

\bibitem{fes85}
 H. Fesefeldt,
 {\em The simulation of hadronic showers: physics and applications},
 Report PITHA-85/02 (1985), RWTH Aachen.

\bibitem{pie08}
   T.\ Pierog and K.\ Werner,
 {\em Muon production in extended air shower simulations},
        Phys.\  Rev.\  Lett.\    {\bf 101}, 171101 (2008).

\bibitem{bat07}
G.\ Battistoni, F.\ Cerutti, A.\ Fass\`o, A.\ Ferrari, S.\ Muraro, J.\ Ranft, 
S.\ Roesler, and  P.~R.~Sala,
 {\em  The FLUKA code: description and benchmarking},
AIP Conf.\ Proc.\  {\bf 896}, 31 (2007).

\bibitem{ble99}
 M.\ Bleicher  {\em et al.}, 
% E.\ Zabrodin, C.\ Spieles, S.\ A.\  Bass, C.\  Ernst,
% S.\ Soff, L.\ Bravina, M.\ Belkacem, H.\ Weber, H.\ St\"ocker, and W.\ Greiner,
 {\em Relativistic hadron hadron collisions in the ultrarelativistic quantum molecular dynamics model},
        J.\  Phys.\  G  {\bf 25}, 1859 (1999).

 \bibitem{ost03}
 S.\ S.\ Ostapchenko,
 {\em Contemporary models of high-energy interactions:
  Present status and perspectives},
  J.\ Phys.\ G {\bf 29}, 831 (2003).

  \bibitem{par11}
    R.\ D.\ Parsons, C.\  Bleve, S.\ S.\  Ostapchenko, and J.\  Knapp,
    {\em   Systematic uncertainties in air shower measurements from 
    high-energy hadronic interaction models},
    Astropart.\ Phys.\   {\bf 34},  832 (2011).

 \bibitem{ost16}
 S.\ Ostapchenko, M.\ Bleicher, T.\ Pierog, and K.\ Werner, 
   {\em Constraining high energy interaction
   mechanisms by studying forward hadron production at the LHC}, 
  Phys.\ Rev.\  D    {\bf 94},   114026 (2016).
	
 \bibitem{ost14}
	S. Ostapchenko, 
 {\em LHC data on inelastic diffraction and uncertainties in the predictions for longitudinal extensive air shower development},
 Phys.~Rev.~D    {\bf  89},  074009 (2014). 


 \bibitem{ost16a}
S.~Ostapchenko and M.~Bleicher, 
{\em Constraining pion interactions at very high energies 
by cosmic ray data},
Phys.\ Rev.\  D    {\bf 93},   051501(R) (2016).

 \bibitem{dre05}  H.\ J.\  Drescher, A.\  Dumitru, and M.\  Strikman,
{\em High-density QCD and cosmic ray air showers},
    Phys.\ Rev.\ Lett.~{\bf 94}, 231801     (2005).

 \bibitem{alb25}
   J.\ Albrecht  {\em et al.},
  {\em   Global tuning of hadronic interaction models with accelerator-based and astroparticle data}, 
  Nat.\ Rev.\ Phys.\  (2025), {\em https://doi.org/10.1038/s42254-025-00897-3}.
%  Road map for the tuning of hadronic interaction models with accelerator-based and astroparticle data},
% arXiv: 2508.21796 [astro-ph.HE].

 \bibitem{kel06}
    S.\ R.\ Kelner, F.\ A.\ Aharonian, and  V.\ V.\  Bugayov,
 {\em    Energy spectra of gamma-rays, electrons and neutrinos produced at proton-proton interactions in the very high energy regime},
 Phys.~Rev.~D    {\bf  74},  034018 (2006). 

 \bibitem{kam06}
   T.\ Kamae, N.\ Karlsson, T.\ Mizuno, T.\ Abe and T.\ Koi,
  {\em        Parameterization of $\gamma$, $e^{\pm}$ and neutrino spectra produced 
  by $p-p$ interaction in astronomical environment},
        Astrophys.\ J.\   {\bf 647}, 692 (2006).

 \bibitem{kos18}
   M.\ Korsmeier, F.\ Donato, and  M.\ Di Mauro,
 {\em  Production cross sections of cosmic antiprotons in the light of new data from the NA61 and LHCb experiments},
Phys.~Rev.~D    {\bf   97},  103019 (2018). 

 \bibitem{kac19}
     M.\ Kachelrie\ss,  I.\ V.\  Moskalenko, and S.\ Ostapchenko,
  {\em 	AAfrag: Interpolation routines for Monte Carlo results on secondary production in proton-proton, proton-nucleus and nucleus-nucleus interactions},
    Comput.\ Phys.\ Commun.\  {\bf 245}, 106846 (2019).

 \bibitem{oru22}
   L.\ Orusa, M.\ Di Mauro, F.\ Donato, and  M.\ Korsmeier,
 {\em New determination of the production cross section for secondary positrons and electrons in the Galaxy},
 Phys.~Rev.~D    {\bf  105},  123021 (2022). 

 \bibitem{oru23}
   L.\ Orusa, M.\ Di Mauro, F.\ Donato, and  M.\ Korsmeier,
 {\em New determination of the production cross section for $\gamma$ rays in the Galaxy},
 Phys.~Rev.~D    {\bf  107},  083031 (2023). 

 \bibitem{oru26}
   L.\ Orusa, M.\ Di Mauro, and F.\ Donato,
 {\em New determination of the neutrino hadronic production cross sections from GeV to beyond PeV energies},
 Phys.~Rev.~D    {\bf   113},   2 (2026). 

	
\end{thebibliography}
\end{document}